\documentclass{emulateapj}
\usepackage{placeins}
\usepackage{graphicx}

\shorttitle{Gap Opening in MHD disks}

\begin{document}

\title{Low mass planets in protoplanetary disks with net vertical magnetic fields:
the Planetary Wake and Gap Opening}

\author{Zhaohuan Zhu \altaffilmark{1},
James M. Stone \altaffilmark{1}, and Roman R. Rafikov \altaffilmark{1}}

\altaffiltext{1}{Department of Astrophysical Sciences, Princeton University, Princeton, NJ, 08544}
\altaffiltext{}{
The movies can be downloaded at http://www.astro.princeton.edu/$\sim$zhzhu/Site/Movies5.html}
\email{zhzhu@astro.princeton.edu}

\newcommand\msun{\rm M_{\odot}}
\newcommand\lsun{\rm L_{\odot}}
\newcommand\msunyr{\rm M_{\odot}\,yr^{-1}}
\newcommand\be{\begin{equation}}
\newcommand\en{\end{equation}}
\newcommand\cm{\rm cm}
\newcommand\kms{\rm{\, km \, s^{-1}}}
\newcommand\K{\rm K}
\newcommand\etal{{\rm et al}.\ }
\newcommand\sd{\partial}
\newcommand\mdot{\rm \dot{M}}
\newcommand\rsun{\rm R_{\odot}}
\newcommand\yr{\rm yr}

\begin{abstract}

We study wakes and gap opening by low mass planets in gaseous 
protoplanetary disks 
threaded by net vertical magnetic fields
which drive magnetohydrodynamical (MHD) turbulence through 
the magnetorotational instabilty (MRI), using three dimensional
simulations in the unstratified local shearing box approximation.
The wakes, which are excited by the planets, are damped by shocks similar to 
the wake damping in inviscid hydrodynamic  (HD) disks.
Angular momentum deposition by shock damping opens gaps 
in both MHD turbulent disks and
inviscid HD disks
even for low mass planets,  in contradiction to the ``thermal
criterion'' for gap opening.  To test the ``viscous criterion'', we compared  gap properties in MRI-turbulent disks to
those in viscous HD disks having the same 
stress, and found that the same mass planet opens a significantly deeper and wider gap 
in net vertical flux MHD disks than in
viscous HD disks. 
 This difference arises due to the efficient magnetic field transport
 into the gap region in MRI disks, leading to a larger effective
$\alpha$ within the gap. Thus, across the gap, the Maxwell stress profile 
is smoother than the gap density profile, and a deeper gap is needed for
 the Maxwell stress gradient to balance the planetary torque density. 
We also confirmed the large excess torque
close to the planet in MHD disks, and found that long-lived density
features (termed zonal flows) produced by the MRI can affect planet migration. 
The comparison with previous results from net toroidal flux/zero flux MHD simulations
 indicates that the magnetic 
field geometry plays an important role in the gap opening process.
Overall, our results suggest that gaps can be commonly produced by low mass planets in 
realistic protoplanetary disks, and caution the use of a constant $\alpha$-viscosity to model
gaps in protoplanetary disks.
\end{abstract}


\section{Introduction}

During the past decade, exoplanet surveys (e.g. $Kepler$) have greatly
advanced our knowledge of the demographics of exoplanets (Howard \etal 2012, Dong \& Zhu 2013), which in
turn sheds light on planet formation.  In the near future, ALMA
will directly probe young planets still embedded in gaseous
protoplanetary disks.  Interpreting such
observations will require a better understanding of how planets
interact with gaseous disks.  In particular, it is well
known that planets can excite density waves in disks, which can lead to
angular momentum exchange between planets and disks and cause planets to migrate
(Goldreich \& Tremaine 1979, Ward 1986, Tanaka \etal 2002, Kley 
\& Nelson 2012, and references therein).  When
these excited waves deposit angular momentum back to the disk, gaps will be opened by planets in disks
(Lin \& Papaloizou 1979, 1993, Artymowicz \& Lubow 1994, Takeuchi
\etal 1996, Bryden \etal 1999, Kley 1999).  Gap opening is 
important for understanding both planet migration and growth.  After
a gap is opened, planet migration slows to the disk viscous timescale
(Lin \& Papaloizou 1986, Nelson \etal 2000), which is normally
referred to as Type II migration.  Slowing down of migration is
important for the survival of planets on the 10$^{6}$ yrs disk life
time.  Moreover, after a gap opens accretion onto the planet is
limited to flow from the circumplanetary disk (Lubow \etal 1999 and
Lubow \& D'Angelo 2006).  The circumplanetary disk 
controls both the final mass of the planet and the formation of
planetary satellites (Ward \& Canup 2010).  A better understanding
of each of these topics motivates the study of gap opening in more
realistic models of protoplanetary disks.

Conventionally, it is thought that two conditions must be fulfilled
simultaneously 
for a planet to open a gap (Lin \& Papaloizou 1993, Bryden \etal
1999).  The first is the ``thermal criterion'', which states that
the planet's Hill radius needs to be larger than the disk scale
height so that the density wave shocks just as it is excited. In
other words, the planet mass needs to be higher than the so called
- local disk thermal mass 
\begin{equation}
M_{th}=\frac{c_{s}^{3}}{G\Omega_{p}}\approx 0.089
M_{J}\left(\frac{c_{s}}{0.6 \rm{km}
s^{-1}}\right)^{3}\left(\frac{M_{\odot}}{M_{*}}\right)^{1/2}\left(\frac{R_{p}}{5
\rm{AU}}\right)^{3/2}\,, 
\end{equation} 
where $M_{J}$ is the mass of Jupiter,
$c_{s}$ is the disk local sound speed,and $R_{p}$ and $\Omega_{p}$ are the
planet's semi-major axis and orbital frequency. 
Recently, this ``thermal criterion'' for gap opening has been
questioned, since
even small amplitude waves can still steepen and
shock in disks after traveling some distance (Goodman \& Rafikov 2001, Rafikov 2002a, Muto \etal
2010, Dong \etal 2011b). This makes possible gap opening
by planets with masses significantly lower than the thermal mass
$M_{th}$, see Rafikov (2002b), Li \etal (2009), Duffell \& MacFadyen (2012).

The second
condition is the ``viscous criterion'', which states that the disk
viscosity must be low enough for the torque from the planet
to overcome the viscous torque. Combined with the ``thermal criterion'', 
this requires (Lin
\& Papaloizou 1993) 
\begin{equation}
\frac{M_{p}}{M_{*}}\gtrsim\frac{40\nu}{R_{p}^{2}\Omega_{p}}\label{eq:viscous}
\end{equation} 
where $\nu$ is the kinematic viscosity of the viscous HD disks.
This ``viscous criterion'' has been analyzed in detail by Crida
\etal (2006) through studying the stream lines and a refined formulae
was suggested.

An obvious difficulty with the 
``viscous criterion'' is that protoplanetary disks are in fact inviscid; angular
momentum transport and accretion are thought to be associated with
MHD turbulence driven by the magnetorotational instability (MRI,
Balbus \& Hawley 1991, 1998 for a review).  In ideal MHD the MRI has
been extensively studied using both local  (Hawley \etal 1995, Stone
\etal 1996, Miller \& Stone 2000, Davis \etal 2010, Guan \& Gammie
2011) and global numerical simulations (Armitage 1998, Hawley  2000,
2001, Fromang \& Nelson 2006, Flock \etal 2011, Beckwith \etal
2011). Some aspects of MRI-driven turbulence can be represented by
an ``$\alpha$'' disk model (Shakura
\& Sunyaev 1973) in a statistical sense (Balbus \& Hawley 1991,1998, Balbus \&
Papaloizou 1999):
\begin{equation}
\alpha=\frac{T_{MRI}}{\rho_{0}c_{s}^{2}}=\alpha_{Re}+\alpha_{Max}=\frac{\langle\langle\rho v_{x}v_{y}'\rangle\rangle}{\rho_{0}c_{s}^{2}}-\frac{\langle\langle B_{x}B_{y}\rangle\rangle}{4\pi\rho_{0}c_{s}^{2}}\,
\label{eq:alpha}
\end{equation}
where  $T_{MRI}$ is the averaged $xy$ component of the total stress
tensor, $\alpha_{Re}$ is the normalized Reynolds
stress, $\alpha_{Max}$ is the normalized Maxwell stress, and
 the double angle brackets $\langle\langle\rangle\rangle$
denotes both time and space averages.
However, there is no guarantee that $\alpha$ is a constant
in disks with large scale density structures (e.g. gaps). Normally, global
simulations (Sorathia \etal 2012) find large $\alpha$ fluctuations at different parts of the disk.

Gap opening by massive planets (much larger than a thermal mass)
in MRI-turbulent disks has been studied by
several authors using net toroidal flux or zero flux MHD simulations 
(Winters et al. 2003, Nelson \& Paploizou 2003, Papaloizou \etal 2004, Uribe \etal 2011). 
Global simulations by Nelson \& Papaloizou (2003) suggest
the gaps in MRI disks are slightly deeper and wider compared with those in viscous disks, while
Winters \etal (2003)  found that gaps are slightly shallower in MRI disks.  Overall, they
found that the shapes of gaps opened by high mass planets
are not very different between MRI and viscous disks. 

Papaloizou \etal (2004) found that both local shearing box MHD simulations
and global MHD simulations show good agreement on the
properties of gaps.  The local approximation enables high resolution
simulations to be carried out.  Furthermore, local shearing box simulations
can maintain a steady state, which is crucial to study planetary
wake and gap properties since some features can only be identified
by averaging over hundreds of orbits.  Although
local simulations lack curvature terms which prohibit studying the differential
torque which controls planet migration, it is still possible to
study the one-sided Lindblad torque and the gap opening process in the local
shearing box approximation as
long as the gap structure is sharper than the disk radius $R$ so
that the curvature  terms are not important. 

The tidal torque felt by the planet 
can be affected by the presence of magnetic fields in disks, which has been studied 
by linear calculations (Terquem 2003) and numerical simulations 
(Fromang \etal 2005, Muto \etal 2008). However, these calculations assume
strong magnetic fields and non-turbulent disks. It is unclear
how MRI turbulence will affect the tidal torque. Recent global MHD
simulations have shown some additional torque close to the planet (Uribe \etal 2011;
Baruteau \etal 2011).  Local shearing box
limit is a good way to test if this additional torque is associated with disk global curvatures. 

Most previous work on gap opening in MHD disks assumed zero net magnetic flux in simulations. 
In this paper, we focus on MHD disks
with a small net vertical magnetic field.  The
reason for choosing a small net field is three-fold. First, recent unstratified
simulations have shown that the turbulent properties depend on both the
numerical resolution and box size with zero net field (Fromang \&
Papaloizou 2007). Second, simulations with net magnetic flux are more
realistic.  We expect that small local patches of the disk will
be threaded by net
vertical fields from molecular clouds, or by toroidal fields generated
by the large scale disk dynamo, or even both (Sorathia \etal 2010). Third, by varying the
net field strength, we can control the strength of the
MRI turbulence, which allows us to systematically study gap opening
in disks with various turbulence stresses. 

Furthermore, previous work
which study gap opening focused on deep gaps induced by  
massive planets (``Jupiter'' mass range). Here 
we are interested in cases where gaps just start to open in order to test 
the ``thermal'' and ``viscous criterion''. This motivates us to  focus on
gap opening by low mass planets (``earth'' mass range).

This paper is organized as follows.  In \S 2, we introduce our
methods.  To provide a benchmark for comparison, we first study gap
opening in inviscid hydrodynamic (HD) disks in \S 3, while planetary wake
properties and gap opening
in MRI-driven turbulent disks is presented in \S 4.  
Gap opening criteria are discussed in \S 5.
After discussions in \S 6, we conclude the paper in \S 7.
 
\section{Method}
\subsection{Basic Equations}

This  study uses Athena (Stone \etal 2008), a higher-order
Godunov scheme for MHD with piecewise parabolic method (PPM) for
spatial reconstruction, the corner transport upwind (CTU) method
for multidimensional integration, and the constrained transport
(CT) to conserve the divergence-free property for magnetic fields
(Gardiner \& Stone 2005, 2008).  The simulations are set up as
isothermal unstratified disks in the local shearing box approximation
(Hawley \etal 1995, Stone \&  Gardiner 2010), in which 
the MHD equations are solved in a reference frame centered at radius $R_{0}$ corotating
with the disk at orbital frequency $\Omega_{0}=\Omega(R_{0})$.
Ignoring curvature terms, the isothermal MHD equations are 
\begin{equation}
\frac{\partial \rho}{\partial t}+\nabla\cdot(\rho \bold{v})=0\,,
\end{equation} 
\begin{equation} \frac{\partial \rho \bold{v}}{\partial
t}+\nabla\cdot(\rho\bold{v}\bold{v}+\rm{T}_{B}+\rm{T}_{v})+\nabla
{\it p}=2qx\rho\Omega_{0}
^{2}\bold{\hat{i}}-2\Omega_{0}\hat{k}\times\rho\bold{v}\,,
\end{equation}
\begin{equation} \frac{\partial \bold{B}}{\partial
t}-\nabla\times(\bold{v}\times\bold{B})=0 
\end{equation}
where $p$ is the gas pressure,
the magnetic stress tensor $\rm{T}_{B}$ is
\begin{equation} \rm{T}_{B}=(B^{2}/2)\rm{I}-\bold{B}\bold{B}\,,
\end{equation}
and the shear parameter $q$ is 
\begin{equation} q=-\frac{1}{2}\frac{d \rm{ln}
\Omega^{2}}{d \rm{ln} r}\,, 
\end{equation} 
so that for Keplerian
flow $q$=3/2.  Note that these equations are written in units
which assume the magnetic permeability $\mu$ =1.
An equation of state for an isothermal gas
$p=c_{s}^{2}\rho$ is used.

Note that, for HD disks, we have included explicit viscosity in the momentum equation, through
the viscous stress tensor $\rm{T}_{v}$
\begin{equation} \rm{T}_{v;ij}=\nu\rho
\left(\partial_{i}v_{j}+\partial_{j}v_{i}-\frac{2}{3}\partial_{k}v_{k}\delta_{ij}\right)\,.\label{eq:rost}
\end{equation} 
where $\nu$ is the kinematic viscosity.  This allows us to study HD models with
viscosity, and compare these to the MHD models which are always inviscid.

An orbital advection scheme (Masset 2000) has been implemented in
Athena (Stone \& Gardiner 2010).  The $y$-direction momentum equation
is split into a linear operator for advection with the background
flow velocity, as well as the MHD equations for the velocity fluctuations.
The orbital advection scheme
significantly accelerates the simulation since the maximum allowed
time step is limited by the magnitude of the velocity fluctuations instead of
the orbital velocity.  It also reduces the inhomogeneous truncation
error produced by differential rotation (Johnson \etal
2008).

Gressel \& Ziegler (2007) have pointed out that the shearing box
boundary conditions can destroy conservation, since the integral
of the fluxes over the two radial faces are not identical to machine precision.
However, in order to model the
density structure (gap formation) in the disk,
it is important that the total mass is conserved
to round-off error. Thus, we have used
the remap scheme similar to Stone \& Gardiner (2010) for the density fluxes.
First, we remap
the radial component of the mass flux at each radial face. Then,
we use the arithmetic average of the radial mass flux computed for
each grid cell at each radial face, and the remapped value of the
radial mass flux from the corresponding grid cell on the opposite
face, to update the density in the cells next to the boundary. This
scheme guarantees the integral of the mass flux is identical at two
boundaries in the $x$-direction.

To simulate the effect of a planet in 3D unstratified shearing box simulations,
a line mass (symmetric in the $z-$direction) is
placed at the center of the box.  This geometry is identical to the symmetry of 2D HD
simulations, but allows MHD turbulence to be modeled in full 3D.
For comparison to 2D simulations we average quantities in the $z-$direction.

The planet potential is smoothed over several grid cells to avoid
small time steps associated with the divergence of a point source.
Following Dong \etal 2011a, we use fourth order smoothing 
\begin{equation}
\Phi_{p}(d)=-GM_{p}\frac{d^2+3R_{s}^{2}/2}{(d^2+R_{s}^{2})^{3/2}}\,,\label{eq:fourthp}
\end{equation}
where $d$ is the
distance to the $z$-axis ($d=\sqrt{x^2+y^2}$) and
$R_{s}$ is the smoothing length.
This potential converges to that of a point mass as $(R_{s}/d)^{4}$ for $d\gg R_{s}$ and 
deviates from the point mass potential by only 1$\%$ at 2.3 $R_{s}$.

Due to the periodic boundary condition, the gradient of the potential has a discontinuity
at the boundary.  Thus we smooth the potential as
\begin{eqnarray}
\Phi_{s,p}(d)&=&\Phi_{p}(d_f)-\sqrt{(\Phi_{p}(d_f)-\Phi_{p}(d))^{2}+G^{2}M_{p}^{2}/R_{so}^{2}}\nonumber\\
&&\,\,\rm{if}\,\it{d}<d_{f}\label{eq:phi}
\end{eqnarray} 
and
\begin{equation}
\Phi_{s,p}(d)=\Phi_{p}(d_{f})-GM_{p}/R_{so}\,\,\,\,\rm{if}\,\it{d}>d_{f}
\end{equation}
where $d_{f}$ is the cut-off radius, beyond which the potential is flat, and $R_{so}$
is a smoothing length at the cut-off radius. For large $R_{so}$, eq. (\ref{eq:phi})
reduces to (\ref{eq:fourthp}). With a finite $R_{so}$ the potential becomes flat smoothly
(the gradient is a continuous function) at $d=R_{so}$. In all our simulations $d_{f}$=7.5 H and
$R_{so}$=50 H, where H is the disk scale height. With this set-up this potential differs from a point
mass potential only by 3.5$\%$ at $d_{f}$=6 H.

Introduction of the planet potential can disturb the background
flow in the disk significantly, and with periodic boundary
conditions this initial disturbance can affect later disk evolution.
Thus, in HD simulations, we linearly ramp up the amplitude
of the planet potential from 0 to 5 orbits, while in MHD simulations,
we linearly ramp up the potential from 5 to 10 orbits after the 
MHD turbulence is established at 5 orbits.

The smoothing length in eq. (\ref{eq:fourthp}) for simulations with
a 0.1, 0.3, and 1 M$_{t}$ planet is chosen as 0.125, 0.18, and 0.269
H respectively, so that the free-fall timescale onto the planet
(defined as Eq. 29 of Dong \etal 2011a) is 0.125 in our code units.
The numerical time step is fixed at 2$\times$10$^{-3}$ ($\sim$60
times smaller than the free fall timescale) even though the orbital advection
scheme allows for a much longer time step. This small time step
allows us to trace the fluid motion around the planet accurately enough to
avoid unsteady fluid motion (Dong \etal 2011a). 
With Athena, Dong \etal (2011a) found, for the case
they studied (with resolution 64 grid points per scale height), that the
numerical time step needs to be at least 140 times smaller than the
free fall timescale onto the planet. While it is clear from physical grounds
the time step must be smaller than
the free fall time near the planet to properly resolve orbital motion of
gas near to the planet, the exact ratio between the time step
and the free fall time can vary between different codes (Kley \etal
2012).
By exploring a larger parameter
space, we have also found that the critical time step is roughly proportional
to the grid size, as one might expect from the CFL condition.
Thus, with a
factor of 2 lower resolution in our simulations (32 grid points per
scale height as discussed below), the critical time step can be 70
times smaller than the free fall timescale. We have run HD
simulations with our choice of parameters and confirmed that all
the runs are stable for 400 orbits with the same boundary set-up in
Dong \etal (2011a).

\subsection{Model Set-up}

We have carried out both 2-D HD (inviscid and viscous)
and 3-D MHD simulations with a 0.1, 0.3, and 1 thermal mass ($M_{th}$)
planet at the box center. The code units are chosen as $c_{s}=1$,
$\Omega=1$, and $\rho_{0}=1$, so that the disk scale height (H) and
the thermal mass ($M_{th}$) are both 1.  The numerical resolution
is 32 grid points per H, and the box size is 16 H $\times$16 H
$\times$ 1 H in most simulations.
The parameters of the
simulations are summarized in Tables 1 and 2.  We divide the simulations into one of the
 three categories. 

1) Inviscid  2-D HD simulations with various box sizes
and a 0.1 $M_{th}$ planet in the disk.  This set of simulations is
designed to test whether linear waves launched by the planet will
steepen into shocks within the simulation domain, and whether
shocks entering from the radially periodic boundary conditions
significantly affect the gap region.  We use these simulations to
determine the ideal box size for further study.

2) 3-D unstratified ideal MHD simulations with and without planets.
Simulations without planets allow us to study
density fluctuations and turbulent stress due only to the MRI. These calculations serve as
a reference for comparison to simulations with planets in order to
distinguish, for example, the density dip due to a
gap opened by a planet from that associated with zonal flows due to the
turbulence (Johansen \etal 2009, Simon \etal 2012). 

MHD simulations with different mass planets 
(0.1, 0.3 and 1 M$_{t}$) 
 have been studied. Two different initial vertical field strengths
 have been applied with $\beta_{0}$=400 and 1600
(the ratio between the initial gas pressure and the magnetic pressure), leading to
disk MHD turbulence with $\alpha$= 0.17 and 0.085.  The initial net vertical
flux is conserved in these simulations.
In ideal MHD, the
wavelength for the fastest growing linear MRI mode is $H/\lambda=0.109
\beta_{0}^{1/2}$ (Hawley \etal 1995). With
$\beta_{0}$=400, and 1600, our vertical box size (H) fits 2 and 4
wavelengths of the fastest growing mode.  With 32 grids per H, the
most unstable wavelength is resolved by 16 and 8 grid cells
respectively.  This is sufficient to ensure the properties of the
MRI-driven turbulence are numerically converged  (Hawley \etal 2011, Simon \etal 2009, 
Sorathia et al. 2012).

3) 2-D viscous HD simulations 
with constant kinematic viscosity that have the same
$R-\phi$ stresses as those in MHD simulations.
The gap structures in these viscous disks are 
compared with the gaps in the MHD simulations from 2) above.
The planet masses are again 0.1, 0.3 and 1 $M_{th}$. 
Note that in viscous disks the $xy$ component of the viscous 
stress tensor
 $T_{v}\sim\nu\rho (\partial v_{y}/\partial x + \partial v_{x}/\partial y)\sim \nu\rho_{0} 1.5 \Omega$. 
Equating $T_{v}$ with the MRI stress $T_{MRI}\sim\alpha\rho_{0}c_{s}^{2}$ leads 
to $\nu=\alpha c_{s}^2/1.5 \Omega$. For an example, in order to compare with  
the MHD run M10B400 which has $\alpha$=0.17, the viscous run
V10 needs to have $\nu$=0.17/1.5=0.113.

\subsection{Boundary Conditions}

In our simulations, we have chosen the periodic boundary condition in both the $x-$ (radial)
and $y-$ (azimuthal) directions.
A periodic boundary condition is natural in the azimuthal direction ($y$ or $\theta$). However,
applying it to the radial direction
 effectively introduces an infinite grid of virtual planets at integer spacings of the radial
box size.
A more natural radial boundary condition would be outflow, or the wave damping condition used by
de Val-Borro et al. (2006).
However, to model MHD turbulence in the local shearing box
periodic boundary conditions 
 must be used at both $x$ and $y$ directions.  Therefore, in order to study the effect of the
$x$-direction periodic boundary condition on our results, and to
choose the optimum domain size, we have calculated three inviscid HD disks (runs I1, I2, I3) having
different radial extents: 8H$\times$16H, 16H$\times$16H, and 32H$\times$16H and a
0.1 M$_{t}$ planet at the center. These simulations
 will be discussed in detail in \S 3, and in the following paragraph we focus on
 the effect of the boundary condition. 

Figure \ref{fig:fig1} shows the disk surface density contours at
400 orbits for all three simulations (runs I1, I2, I3).  The effect of the $x$-direction
periodic boundary condition is most apparent in the smallest box
(8 H $\times$ 16 H). The wakes from nearby ``virtual'' planets
overlap with the gap region in the box. Multiple stripes appear in
this box instead of two clean gaps, and the edge of the gap is close
to the $x$-direction boundaries.  While, the biggest box (32 H$\times$
16 H) has the cleanest gap, the intermediate-sized domain (16 H
$\times$ 16 H) resolves the two gap structure well, with the two
gaps well within the simulation domain. Furthermore, the wakes from
nearby planets are much weaker than the wake from the planet at
the center. Thus, as the best compromise between cost and accuracy,
we have chosen 16 H $\times$ 16 H as our box size for most other
simulations. We note that in both viscous and MHD models presented
later, the wakes from ``virtual'' planets are a lot weaker than
those in the inviscid HD model presented here since the wakes are
quickly damped either by viscous stress or the MRI turbulence (\S
4).  The box size used in this study is therefore even more suitable
in these cases.

\section{Gap opening in inviscid HD Disks}
\begin{figure}
\epsscale{1.0} \plotone{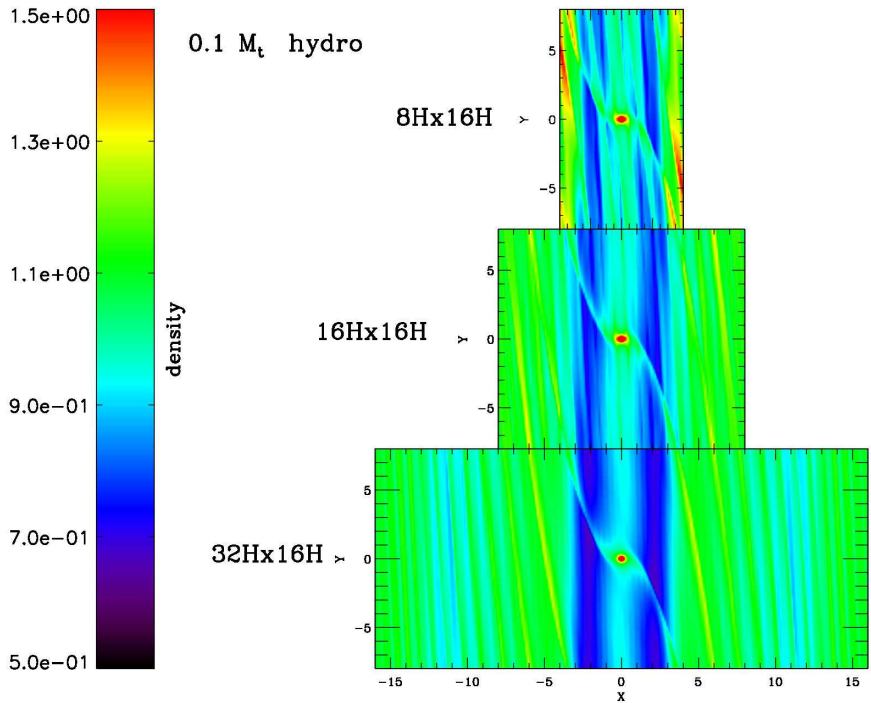} \caption{The surface density in 2-D inviscid HD  disks
with a 0.1 thermal
mass planet at the center. Simulations with different box sizes (cases I1, I2, I3)
have been shown. The snapshots are taken at 400 orbits. A cleaner gap
is seen in a bigger box and the effect of virtual planets from neighbor boxes  diminishes.  The 16 H$\times$16 H box shows the gaps clearly and 
is used for most runs in the paper. The simulations also demonstrate that low mass planets (less
than the thermal mass)
can open gaps in inviscid HD disks.} \label{fig:fig1}
\end{figure}

Figure \ref{fig:fig1} also questions the ``thermal criterion'' for gap opening in
the inviscid fluid limit. Gaps are opened by a planet with mass significantly
smaller than the ``thermal mass''. The objection to the ``thermal criterion'' was first suggested by
Goodman \& Rafikov (2001), stating that 
linear waves from a low mass planet ($M_{p}<M_{th}$) steepen
to shocks after they propagate for a distance
\begin{equation}
|x_{sh}|\approx 0.93 \left(\frac{\gamma+1}{12/5}\frac{M_{p}}{M_{th}}\right)^{-2/5} H\,.\label{eq:eqshock}
\end{equation} 
where $\gamma$ is the adiabatic index. Shocks efficiently 
transfer the wave
angular momentum to the background flow,
and two gaps should gradually open
at the distance of $|x_{sh}|$ on each side of the planet. Thus
the thermal mass $M_{th}$ is not a gap opening threshold in this
nonlinear wave propagation theory.  For a $M_{p}=0.1
M_{th}$ planet in an isothermal disk ($\gamma$=1), the waves shock at the
distance $|x_{sh}|\sim 2.5 H$.

In Figure \ref{fig:fig1}, we clearly see gap opening 
for a 0.1 $M_{th}$ planet in an inviscid HD disk
(runs I1, I2, I3 ).  The gaps are indeed
deepest at $|x|\sim x_{sh}\sim2 H$. The gaps are more prominent with bigger boxes
since they are less affected by the boundary conditions,
as discussed in \S 2.3.

\begin{figure}
\epsscale{1.0} \plotone{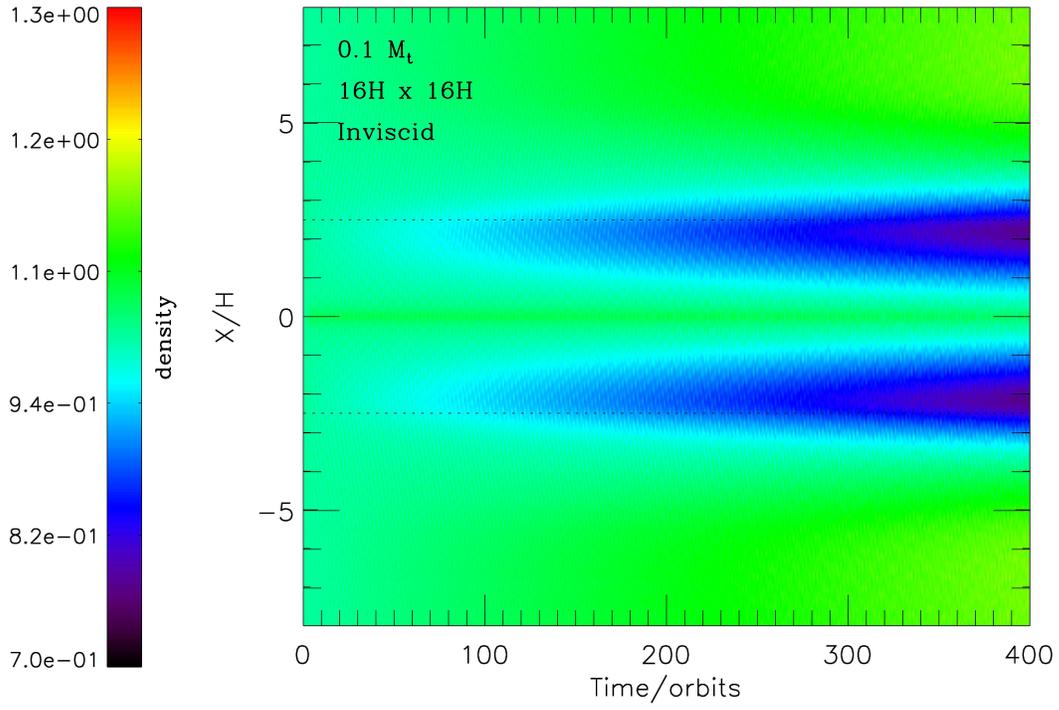} \caption{The space-time plot for the disk surface density ($y$-direction averaged)
 with respect to $x$ in the inviscid HD  disk (case I2) having a 0.1 thermal mass planet at the center.
 $X$ axis is in unit of orbits. Two gaps are gradually opened adjacent to the planet  and they get deeper with time. The two dotted lines are where density waves excited by a 0.1 thermal mass planet become shocks.
} \label{fig:fig2}
\end{figure}

Figure \ref{fig:fig2} shows the space time plot
for the disk surface density in run I2 (the same simulation as 
in the middle panel of
Figure \ref{fig:fig1}). The $x$-axis is the
time in units of the orbital period after the planet is inserted while
the $y$-axis is the $y$-direction (azimuthal) averaged disk surface
density across the gap.  We can see 
that two gaps
are gradually opened starting at $|x|\sim2 H$, getting deeper
and wider with time. For a long time (hundreds of orbits) a ribbon of 
material is maintained between the gaps, in agreement with the 
predictions of Rafikov (2002b). This structure appears because 
density waves do not dissipate in the coorbital region in an 
inviscid case, and thus angular momentum is not transferred to 
the disk, leaving the fluid between the gaps essentially 
unperturbed. The 
symmetry of the two-gaps structure in our simulations is due to 
the shearing box geometry, and should be broken in global 
simulations, as well as if planet is migrating through the disk. 

No steady state is reached, since in an inviscid HD  
disk, there is no torque to balance the torque from shock 
dissipation. We find that
gaps continue to widen and deepen until some instability, possibly 
the Rossby wave instability (Lovelace \& Hohlfeld 1978;
Lovelace et al. 1999; Li \etal 2000; de Val-Borro 2007; Lin \&
Papaloizou 2010), or Rayleigh instability sets in and destroys
the ribbon of gas coorbital with the planet. This should finally 
result in a single gap in a disk, but reaching this state can 
take long time for a low mass planet in an inviscid disk. 

The timescale to open a factor of 2 deep gap is $\sim$400 orbits in our
local simulations. However, to derive a timescale for gap opening
in a realistic disk based on our local shearing box simulations,
the time in our simulations needs to be multiplied by $2\pi R$/$16
H$, where R is the distance from the central star to the planet and
$16 H$ is our box size in $y$-direction. This is simply due to the
fact that the angular momentum injected by the planet to remove all the material
per unit radial distance is proportional to the box extent in the
$y$-direction (which should be $2\pi R$ instead of $16 H$), while
the torque excited by the planet and the amount of shock dissipation per unit radial
distance are independent on the box sizes \footnote{The box needs to be 
larger than $\sim 4 H\times 4 H$ so that most of the planetary torque has been excited (Goldreich \& Tremaine 1979).}. 
 Thus, for a global disk with H/R=0.1, the timescale
to open a gap by a 0.1 $M_{th}$ planet is estimated to be 400$\times
2\pi R/16 H$$\sim$ 1,500 orbits.

\section{Gap opening in MRI disks with net vertical flux}
In this section, we will first study the wake properties and wake damping mechanism 
in MHD
turbulent disks
 using simulations with low mass planets 
(e.g. M01B400 and M03B400, \S 4.1). Then we will compare the gap properties between
the MHD turbulent disks and the viscous
disks using simulations with more massive planets (e.g. M10B400, M10B1600, \S 4.2).

\begin{figure}
\epsscale{1.0} \plotone{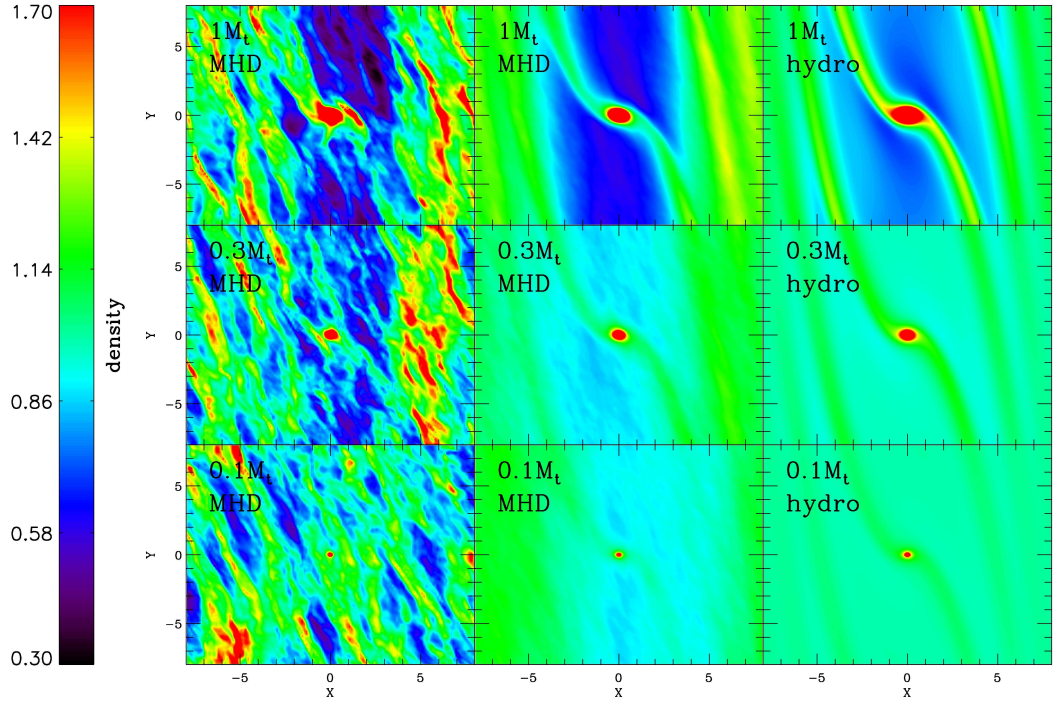} \caption{The $z$-direction averaged disk surface densities for MRI turbulent disks
(left and middle panels, M10B400, M03B400, M01B400 from top to bottom) and viscous disks (2D) having the same
equivalent stress $\alpha$=0.17 as the MRI disks (right panels, V10, V03, V01 from top to bottom). The planets at the box center
have different masses: 
1 thermal mass in the upper panels, 0.3 thermal mass in the middle panels and
0.1 thermal mass in the lower panels. The left panels are the snapshots
of the MHD simulations at a given time while the middle panels are from the same simulations but
 averaged over hundreds of orbits.
Comparing the upper middle and right panels, we can see the gap is deeper
in the MHD case.} \label{fig:fig3}
\end{figure}

The disk surface densities from these simulations 
are shown 
in Figure \ref{fig:fig3} for planet masses of
1 M$_{t}$, 0.3 M$_{t}$, and 0.1 M$_{t}$ (from top to bottom).
The left column shows a snapshot in
the MHD simulations (runs M10B400, M03B400, and M01B400) 
at 70 orbits.  The turbulent  nature of the MRI
disk can be clearly seen. Density waves excited by the MRI turbulence
(Heinemann \& Papaloizou 2009) are present across the whole simulation
domain.  The standard
deviation for density fluctuations relative to the background density due to the MRI turbulence alone is as large
as $\sim$0.3.  With such large density fluctuations, both the gap
structure and planetary wake are hard to discern.  In order to extract
the gap and planetary wake from MRI turbulence, we have averaged
the disk surface density every orbit
from 80 orbits until the end of
the calculation (shown in Table 2).  These averaged surface densities are shown in
the middle column of panels, where both the gap structure and planetary wakes are clearly evident.  By comparing to gaps in
viscous HD disks (runs V10, V03, V01, the right column of panels), it is clear that 
the gaps in MRI disks
are deeper, which will be further explored in \S 4.2.

\begin{figure}
\epsscale{1.0} \plotone{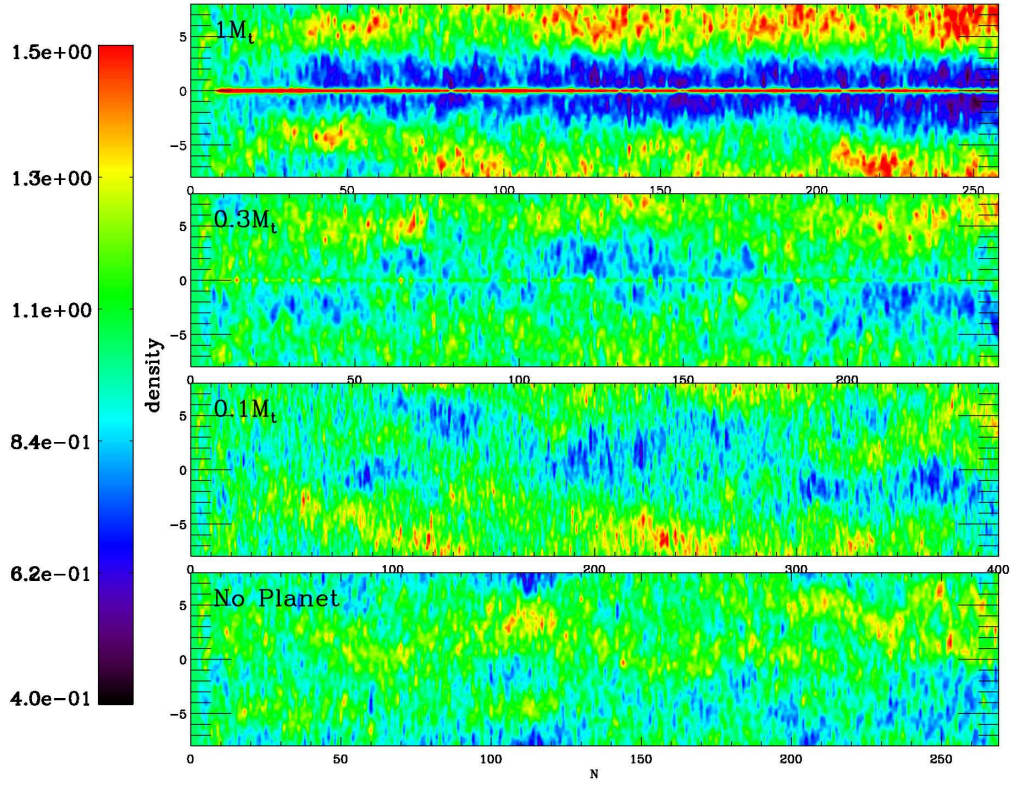} \caption{The space-time plots of the $yz$-direction averaged disk surface densities for M10B400, M03B400, 
M01B400,  and B400 (top to bottom panels). A clear gap is opened in the M10B400, while density structures in lower mass cases 
(asymmetric with respect to $x=0$)  resemble zonal flow 
from B400 (the bottom panel).} \label{fig:fig8}
\end{figure}

It is important to note that MRI-turbulent disks can produce large scale density
structures that persist over tens to hundreds of orbits, called
zonal flows (Johansen \etal 2009, Simon \etal 2012), even without the presence of
a planet.  To distinguish
the gap opened by the planet from zonal flow, we show the
space-time plot for the y-averaged disk surface density for runs
M10B400, M03B400, M01B400 and B400
in Figure  \ref{fig:fig8}. The magnitude of density fluctuations produced by zonal flow can be seen
in the bottom panel for the MRI disk with no planet case(B400).  Clearly, the
density dip in the 1 M$_{t}$ planet run (M10B400, the top panel) is much
larger in magnitude than the zonal flow, and the position of the gap is close to
the planet. Thus we can confidently say that it is indeed a gap induced
by the planet in the M10B400 case. Before we study the gap opening process
in MHD disks (\S 4.2), we first need to understand 
how waves deposit angular momentum in MRI turbulent disks (\S 4.1).

\subsection{Wake Properties}
\subsubsection{Identifying Wakes}
The damping mechanism of propagating waves determines
how angular momentum is deposited into the disk and furthermore 
the gap opening process. 
Various damping mechanisms have been proposed, including 
viscous damping (Takeuchi \etal 1996) in viscous disks
and shock damping (Goodman \& Rafikov 2001) in inviscid HD disks.  
In this work we will investigate the damping mechanism in the high Reynolds number
MHD turbulence expected in real protoplanetary disks.

\begin{figure}
\epsscale{1.0} \plotone{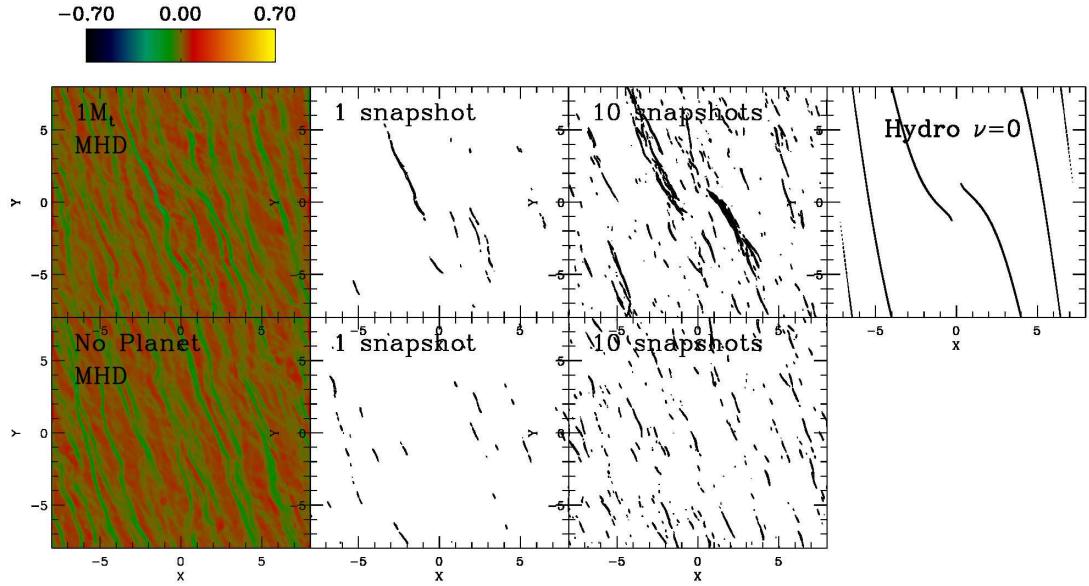} \caption{The $z$-direction averaged $(\Delta x/ c_{s})\nabla\cdot v$ of MRI disks with
a 1 thermal mass planet at the center (upper left three panels, M10B400) and the MRI disk without planets (lower panels, B400). The leftmost panels show 
the color contours of  $(\Delta x/ c_{s})\nabla\cdot v$ at a given time. The second to left panels outline the contours with values smaller than -0.2, which identify
the shock regions in disks. The second to right panels overlap these outlines in snapshots
from 80 to 130 orbits  with 5 orbits interval. The upper right panel shows the same 
outline for an inviscid  HD disk (I10inv). The planetary wake in the MHD case has similar shock strength as the wake in the inviscid HD case, thus
in MRI disks the waves are dissipated by shocks.
However, since the shock position varies with time, the averaged wake density profiles over hundreds of orbits are a lot smoother as shown in Figure \ref{fig:fig5}. } \label{fig:fig4}
\end{figure}

To identify the shock region, in Figure \ref{fig:fig4} we have
plotted the vertical average of the velocity divergence $\eta \equiv
(\Delta x/ c_{s})\nabla\cdot v$ (using the sound speed and grid
spacing to normalize the velocity and length scales respectively)
for both MRI disk (B400, bottom panels) and MRI+planet disk (M10B400,
top panels).  In our simulations, we find $\eta < -0.2$ is a good criterion
for identifying strong shocks.

The leftmost panels of Figure \ref{fig:fig4} show the contours of
$\eta$ at 80 orbits.
The next column of panels outline the region where
$\eta<$-0.2, while the next column after that
show these outlines in every snapshots from 80 to
130 orbits using a 5 orbit interval.  Even in the case of no
planets (B400), as shown in the bottom panels, sharp density features are
present in the MRI-driven turbulent disks. These shocks in MRI
disks are consistent with structures explored in Heinemann \&
Papaloizou (2012). However, due to the random nature of the turbulence,
these shocks appear uniformly in the disk (the lower right panel).
When a 1 $M_{th}$ planet is present in the disk (M10B400), as shown in the
upper panels of Figure \ref{fig:fig4}, it excites strong shocks
around the planet, and  $\eta$ can be as low
as -0.7.  For comparison, a planet in a 2-D inviscid  HD disk (I10inv)
also excites a strong shock
(as shown in the upper rightmost panel), and the smallest  $\eta$  is also -0.7.
On the other hand,  in the equivalent viscous disk simulation (V10), $\eta$ of the wake is only -0.066,
one order of magnitude smaller than that in both the MHD and inviscid
HD  cases.  Thus, we conclude that shock damping is the main wake dissipation
mechanism in MRI disks, which is
significantly different from viscous damping with a
kinematic viscosity.

As shown in the upper panels of
Figure \ref{fig:fig4}, the wake's shock
 position in MRI-turbulent disks varies with time due to the
disk turbulence (the second to right panel), which is different from the static shock in inviscid HD case (the rightmost panel).  
The random fluctuations in the shock
position appear to spread it out over a finite distance, even though
at any instant in time the shock profile is sharp.  As we will show
below, the time average density profile of the planetary wake is
surprisingly similar to the wake in a viscous disk.

\subsubsection{Averaged Wake Density Profile}
The one-armed planetary wake, which is located at
\begin{equation}
y_{t}\approx -sgn(x)\frac{3x^{2}}{4H}\,,\label{eq:eqyt}
\end{equation}
in the local limit (Goodman \& Rafikov 2001), is due to
the coherent interference (Ogilvie 
\& Lubow 2002) of all the modes excited by the planet
(perturbed quantities $\propto exp[i(m\phi-\omega t)+k_{R}R)]$),
 each satisfying the WKB dispersion relationship
 without self-gravity
\begin{equation}
m^{2}[\Omega(R)-\Omega_{p}]^{2}=\kappa^{2}+k_{R}^{2}c_{s}^{2}\,,\label{eq:eqdis}
\end{equation}
where $\kappa$ is the epicyclic frequency ($\kappa=\Omega$ for a
Keplerian disk), $k_{R}$ is the radial wavenumber in the WKB limit,
and $\Omega_{p}=\Omega(R_{p})$ is the pattern speed $\omega/m$ in the inertial
frame. 

Since this one-armed wake is the ``signpost'' of planet-disk
interaction, it is worth studying its structure in detail.  The
density profiles across the wake in the inviscid HD  disk have
been studied by Goodman \& Rafikov (2001), Dong \etal (2011a)
and Rafikov \& Petrovich (2012).
In this work, we will examine the density profiles of the wake in
MRI-driven turbulent disks.  In a single snapshot, the planetary wake is
significantly disturbed by MHD turbulence (as shown in \S4.1.1
above). Thus we average the disk surface density at every time
step over 100 orbits for the planetary wake to stand out.  

In order to compare the wake at different $x$ positions in different simulations,
we need to apply some scaling procedures as in Goodman \& Rafikov (2001).
For HD simulations, the wake 
density profiles are measured
along azimuthal cuts through the disk at $x$=0.5, 1, 2, 4 H, and then we shift the
derived one dimensional density profiles in $y$ by $y_{t}$ given in
Eq. (\ref{eq:eqyt}). We then scale the density to a dimensionless
quantity as follows (Goodman \& Rafikov 2001): first,
the mean azimuthal density at $x$ is subtracted
to derive $\delta \Sigma$. Since $\delta \Sigma$ increases with x
in the linear regime due to the angular momentum flux conservation,
we normalize $\delta \Sigma$ by x$^{1/2}$. Finally we scale $\delta
\Sigma$ by $\Sigma(x)(M_{p}/M_{th})$, where $\Sigma(x)$ is the 
$y$-direction averaged surface density along $x$.  In the linear
theory, with this scaling,  the density profile should be independent
of M$_{p}$ at a fixed separation $x$, and the magnitude of the wake density
profile should remain constant with increasing $|x|$ (until
the wake shocks).

For MHD simulations, the above scaling procedure is similar but the
scaling speed now is  the fast magnetosonic
speed $\sqrt{c_{s}^{2}+v_{A}^{2}}$, where $v_{A}$ is the Alfven
speed, instead of the sound speed ($c_{s}$) as in HD cases (Terquem 2003).
We find that the MRI saturated states in these
simulations have $\beta\approx$4, thus the fast magnetosonic speed
is  1.22 c$_{s}$ and the new length scale is 1.22 H. 
Therefore we make the density cuts at x=($\pm$)0.61, 1.22, 2.44,
4.88 H. Then we shift the one dimensional density profile
by
\begin{equation}
y_{t}=-sgn(x)\frac{3x^{2}}{4H}\frac{\sqrt{c_{s}^{2}+v_{A}^{2}}}{c_{s}}\,,\label{eq:eqyt2}
\end{equation}
similar to Papaloizou \etal (2004). Finally we scale $\delta \Sigma$
by $\Sigma_{0}(M_{p}/M_{th}')$ where
$M_{th}'=(c_{s}^{2}+v_{A}^{2})^{3/2}/(G\Omega)$ is a new thermal
mass defined via the fast magnetosonic speed.  Since the resulting density
profiles have a slight asymmetry due to zonal flow in the disk, 
we average the density profiles from both the positive and
negative x sides of the disk.  

\begin{figure}
\epsscale{1.0} \plotone{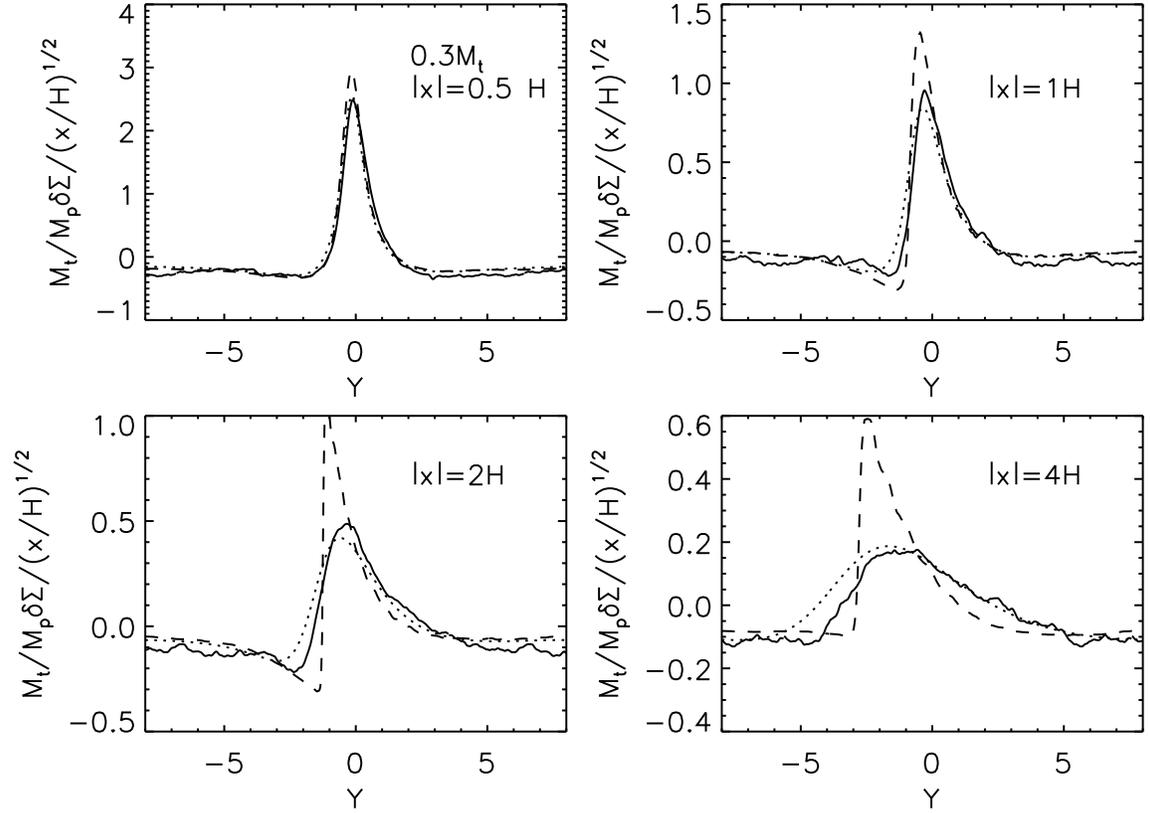} \caption{The (time and $z$-direction averaged) surface density profiles along y-direction at 
x=0.5,1,2,4 scale heights away from the planet in inviscid HD  (I03inv, dashed curves), viscous (V03, dotted curves) and 
MHD simulations (M03B400, solid curves). 
The solid curves are scaled with the fast magnetosonic speed 
($\sqrt{c_{s}^{2}+v_{A}^{2}}$, detailed in \S 4.1.2). A remarkably good agreement is found between the averaged density profiles of
MHD cases and those of viscous HD cases, even though the wave is dissipated by shocks in MHD cases (Fig. \ref{fig:fig4}). 
} \label{fig:fig5}
\end{figure}

The resulting wake density profiles from
a 0.3 M$_{t}$ planet in disks are shown in Figure
\ref{fig:fig5}. Solid, dotted and dashed curves are from MHD simulations (M03B400),
viscous HD (V03), and inviscid HD simulations (I03inv) respectively.
Remarkably, with the new scaling Eq. (\ref{eq:eqyt2}),
the averaged density profiles from the MHD case are quite  similar to the density
profiles in the viscous disk with the same stress.
The sharp density profiles in the inviscid HD  disk (I03inv)
indicate the formation of the strong shock, while the smooth profiles in the viscous
case (V03) indicate the viscous damping.  However,
the smooth density profiles in the MHD case suggest that, although the damping mechanism
is the shock damping (\S 4.1), the temporal fluctuation of the shock
position in an MRI-turbulent disk spreads the wake and, in a time-averaging point of view,
the shocks in turbulence have similar effects as the viscosity on the averaged wake density profiles.

\subsubsection{Averaged Torque}

Using the time averaged surface densities $\langle \Sigma \rangle$
from MHD simulations, we can calculate the time averaged torque
density $\langle dT/dx \rangle$ from the planet by integrating
$\langle \Sigma \rangle \partial\phi/\partial y$ over the y-direction
\begin{equation}
\langle\frac{dT}{dx}\rangle=-\int \langle \Sigma \rangle \frac{\partial \phi}{\partial y}dy\,.\label{eq:torqued}
\end{equation}
where the cylindrical region centered on the planet
 within the planet's Hill radius is excluded for the torque calculation
since this region should be bound to the planet and co-moving with the planet.
Note that we can calculate the torque density by only integrating
$\langle \Sigma \rangle$ over y direction instead of integrating
$\langle \rho \rangle$ over both y and z directions, since the
planetary potential is independent of the z-direction in our
calculations. 

\begin{figure}
\epsscale{1.0} \plotone{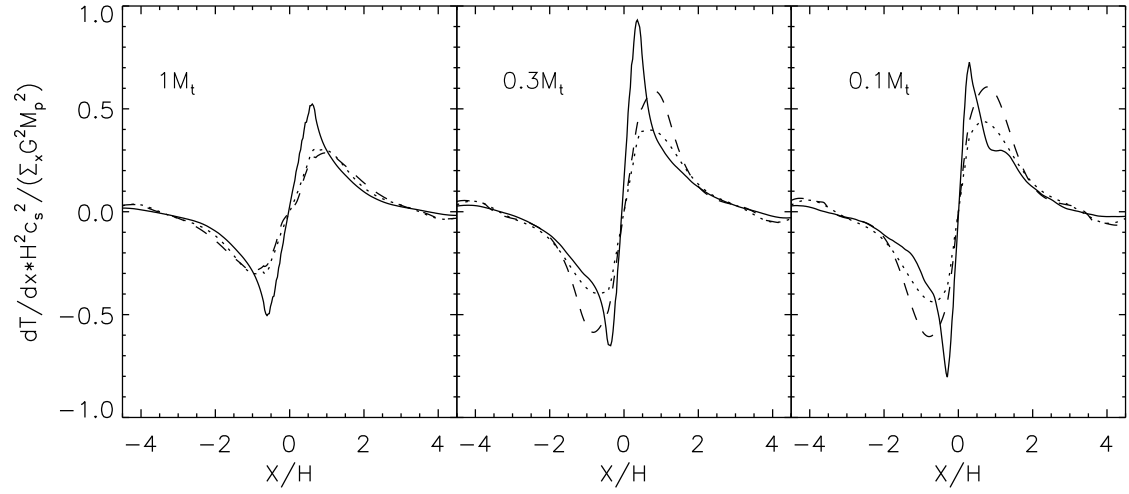} \caption{The time averaged torque densities excited by a 1 M$_{t}$, 0.3 M$_{t}$, 0.1 M$_{t}$
planet in the disk (from left to right). The solid curves are from MHD simulations  (M10B400, M03B400, M01B400). 
 The dotted curves are from viscous HD runs (V10, V03, V01) while the dashed curves are from inviscid HD runs (I10inv, I03inv, I01inv). 
 Torque densities in MRI disks
have peaks closer to the planet. 
} \label{fig:torquet}
\end{figure}

The averaged torque densities for M10B400, M03B400, and M01B400 are shown 
in Figure \ref{fig:torquet} as the solid curves. By comparison, 
the dashed curves are from inviscid HD  simulations (I10inv, I03inv, I01inv), while the
dotted curves are from viscous HD simulations (V10, V03, V01). 
As well known, the torque
densities
for inviscid HD  disks (dashed curves) have peaks around 1 H,
due to the torque cutoff (Goldreich \& Tremaine 1980,
Artymowicz 1993, Rafikov \& Petrovich 2012).
In contrast to both the inviscid and viscous HD cases, the
averaged torque densities from MRI disks have larger peak values
at smaller $|x|$.
More discussions are in \S 7.1.

\subsection{Deeper and Wider Gaps}

As shown in Figure \ref{fig:fig3}, the gaps in MHD-turbulent disks with
net vertical flux (M10B400)
are significantly deeper and wider than the gaps in viscous disks with the
same stress (V10).  To be more quantitative, we average the disk surface
densities from MHD simulations M10B400 \& M10B1600 over both time, and the $y-$ and
$z-$directions to derive the surface density profile along $x$, 
and compare them with those from viscous simulations V10 \& V10sv,
as shown in Figure \ref{fig:fig12}. Due to the large mass concentration in the
circumplanetary region around the planet,
the region within $|y|<H$ is removed for the averaging. The
solid curves are for MHD simulations, while the dotted curves are for viscous HD
simulations.
For run M10B400 (solid black curves), the ratio
between the smallest surface density within the gap and the surface
density at the box edge is $\sim 50\%$, while this ratio is $\sim
80\%$ for the viscous run V10 (dotted black curves). With smaller stresses,
the gaps in M10B1600 (solid grey curves) and V10sv (dotted grey curves) are both deep, but
the gap in the MHD case M10B1600 is still significantly
deeper and wider than the corresponding viscous case V10sv. 

In order to understand why the gaps are significantly deeper and wider in these MHD cases, we note
that the steady-state gap shape is determined by the balance 
between angular momentum deposited into the disk and the 
stress from MRI turbulence (or viscosity). For viscous HD disks, this balance can be expressed as
\begin{equation}
-\int\rho\frac{\partial \phi}{\partial y}dydz-\frac{d\int(\rho v_{x}v_{y})dydz}{dx}=\frac{d\int-\nu\rho(\partial_{x}v_{y}+\partial_{y}v_{x})dydz}{dx}\,.\label{eq:eqstre}
\end{equation}
while for MHD disks, it is
\begin{eqnarray}
-\int\rho\frac{\partial \phi}{\partial y}dydz-\frac{d\int(\rho v_{x}v_{y})dydz\vert_{planet}}{dx}&=&\nonumber\\
\frac{d\int(\rho v_{x}v_{y}\vert_{MRI}-B_{x}B_{y}/4\pi)dydz}{dx}&&\label{eq:eqstre2}
\end{eqnarray}
Note that we separate the Reynolds stress in Eq. (\ref{eq:eqstre2}) into two parts:
the portion excited by the planet and that excited by MRI turbulence.

Next, we integrate the above
equations along $x$ starting from the planet position $x$=0, and then
perform an ensemble average for both time and the $z$-direction. For the
viscous HD disk this gives
\begin{eqnarray}
-\int\langle\frac{dT}{dx}\rangle dx-\int\langle\langle\rho v_{x}v_{y}\rangle\rangle dy\bigg\vert_{0}^{x}&=&\nonumber\\
-\int\langle\langle\nu\rho(\partial_{x}v_{y}+\partial_{y}v_{x})\rangle\rangle dy \bigg\vert_{0}^{x}&&\,.\label{eq:eqstre3}
\end{eqnarray}
while for the MHD disk it gives
\begin{eqnarray}
-\int\langle\frac{dT}{dx}\rangle dx-\int\langle\langle\rho v_{x}v_{y}\rangle\rangle_{planet}dy\bigg\vert_{0}^{x}&=&\nonumber\\
\int\langle\langle\rho v_{x}v_{y}\vert_{MRI}-B_{x}B_{y}/4\pi\rangle\rangle dy\bigg\vert_{0}^{x}&&\label{eq:eqstre4}
\end{eqnarray}
For the disk region from 0 to x, the left-hand sides of Eq. (\ref{eq:eqstre3}) and (\ref{eq:eqstre4})
represent this region's net gain of angular momentum from the waves excited by the
planet. They consist of the integrated torque exerted by the planet
(the first term), minus the angular momentum flux of the waves leaving 
this region (the second term).  The right-hand sides of Eq. (\ref{eq:eqstre3})
and (\ref{eq:eqstre4}) represent the net flux of angular momentum
into this region carried by viscosity, or by MHD stresses.   A steady gap shape is
achieved when the two sides of each equation balance.  

We integrated the torque density from the same mass planet
for both MHD and viscous runs (as derived in \S 4.1.3), and found that the integrated torques (the first
terms on the left-hand singes of  Eqs.  
\ref{eq:eqstre3} and \ref{eq:eqstre4}) are close to each other within  20$\%$ difference.
Considering the torque density normally is much larger than the planet induced
Reynolds stress, the left hand sides of Eqs.  
(\ref{eq:eqstre3}) and (\ref{eq:eqstre4}) are similar.
This suggests that the rate of deposition of angular momentum from the planet to the disk
is similar in both MHD and viscous HD disk simulations, and therefore the differences in
gap depth between these two cases can 
only be due to the stress gradient along $x$ (the right hand sides of Eqs 
(\ref{eq:eqstre3}) and (\ref{eq:eqstre4})) as shown below.

\subsubsection{Viscous versus Maxwell stress}

\begin{figure}
\epsscale{1.0} \plotone{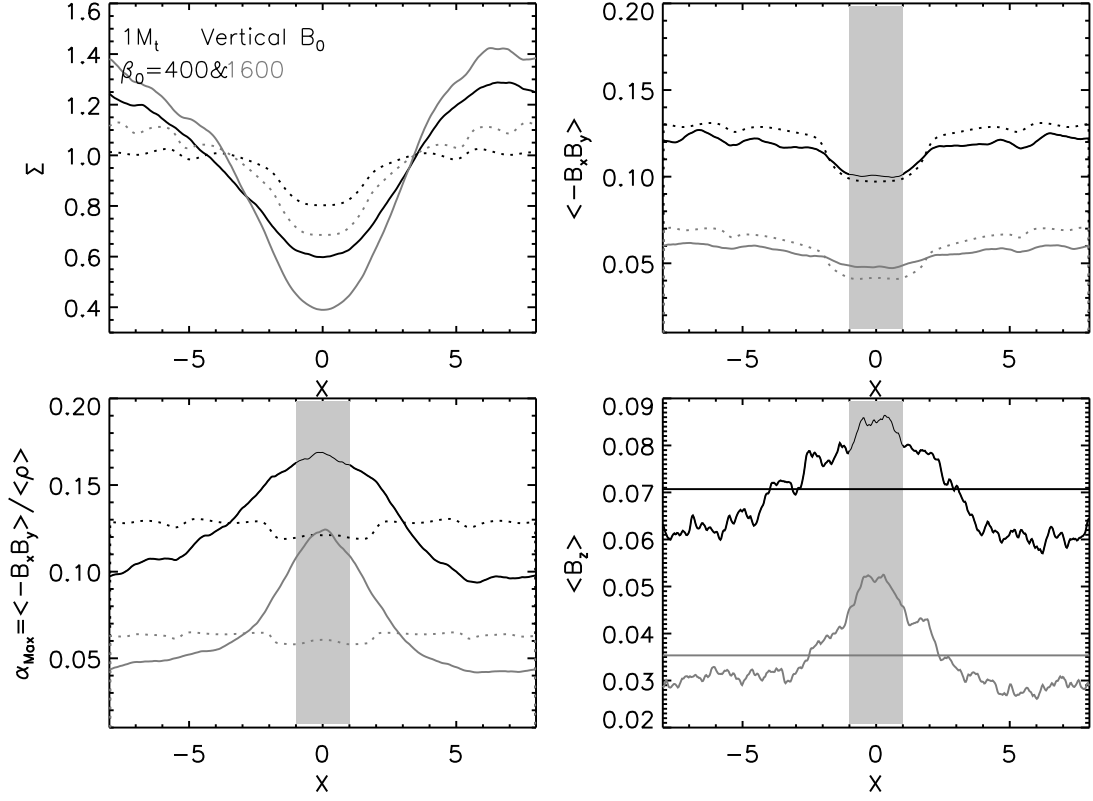} \caption{ The $yz$-direction and time averaged disk
surface densities (upper left panel), Maxwell Stress (upper right panel), equivalent $\alpha_{Max}$ (lower left panel), and net vertical flux (lower right panel) for vertical net flux MHD simulations with a 1 thermal mass planet at the box center.
Regions closing to the planet ($|y|<H$) are masked out for the averaging. $|x|<H$ is shown in
the upper right and lower panels as the shaded region.
 Two different initial field strengths have been applied (M10B400(dark curves) and M10B1600 (light
curves)). The equivalent viscous cases(V10sv, V10st) are over-plotted as the dotted curves. In order to be compared
with the Maxwell stress and $\alpha_{Max}$, the viscous stress ($T_{xy}$) and $\alpha$ are multiplied by 3/4
in these plots. 
Clearly, the gaps are significantly deeper in MHD cases than the viscous cases since in MHD cases the stress is flat compared to the density, or, in other words, $\alpha$ increases towards the gap region. In the lower right panel, the initial vertical
field strengths are plotted as the flat solid lines. Net magnetic fields are concentrated
 in the gap region.
} \label{fig:fig12}
\end{figure}

The averaged viscous stresses for the viscous disks (V10, V10sv) and the Maxwell stresses for
the MHD disks (M10B400, M10B1600) are shown in the upper right panel of Figure \ref{fig:fig12}.
Only $xy$ components of the stresses are shown. By comparing with the density profiles 
in the upper left panel, we see that the viscous
stresses are almost proportional to the densities in the viscous disks (dotted curves). 
This correlation is expected since  the
shear in the $y$ direction dominates the stress and the stress can
be approximated by $1.5\nu\rho\Omega$. $\nu$ and $\Omega$ are both
constant in viscous disks. However, as shown as the solid curves,  in MHD disks,
 the Maxwell stresses across the gaps are significantly smoother than the density
changes.  Since
the gap shape is not determined by the stress itself but by its
gradient along $x$ (Eq. \ref{eq:eqstre2}), this smoother stress with respect
to $\rho$ across the gap
means that to maintain the same stress gradient to balance the planetary torque
the gap needs to be deeper in the MHD case.

\begin{figure}
\epsscale{1.0} \plotone{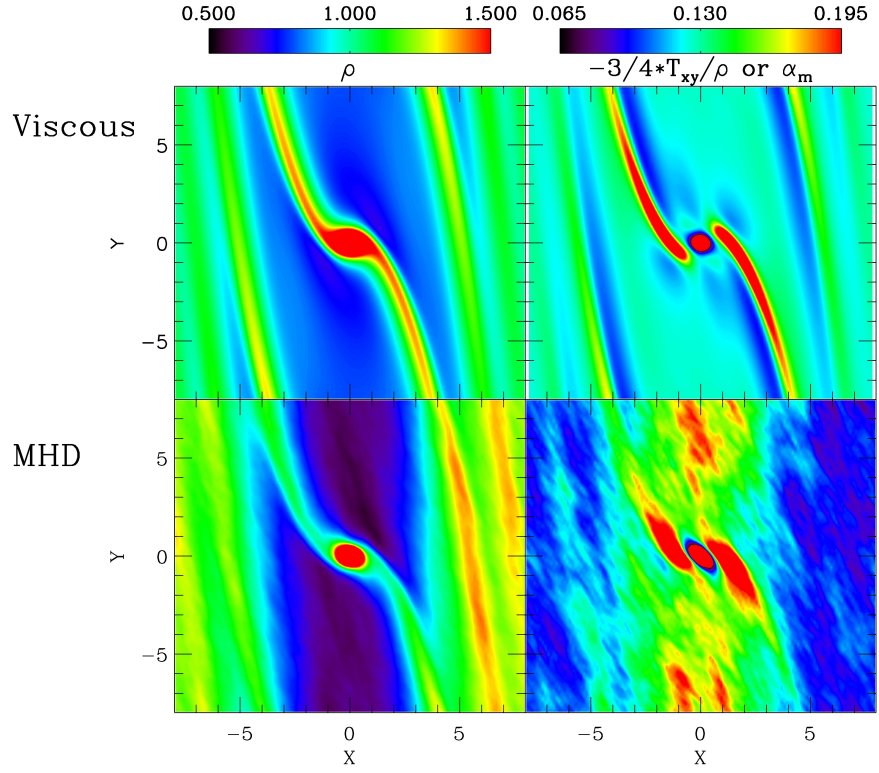} \caption{The disk surface density (left panels),
and the ratio between
the stress and density ($\alpha$, right panels) for viscous (V10, upper panels) and MHD (M10B400, lower panels) cases. 
For the MHD case, $xy$ component of the stress is the Maxwell stress. 
For the viscous case, the viscous stress ($T_{xy}$) is multiplied by 3/4  to be compared
with the Maxwell stress.
Compared with the viscous case,
the stress-density ratio ($\alpha$) in the MHD case is higher in the gap.} \label{fig:fig7}
\end{figure}

Along the $x$ direction, the smoother profiles of the Maxwell stress compared with the density can also be characterized by dividing
the stress with density, or equivalently calculating $\alpha$
parameter (Eq. \ref{eq:alpha}). For the viscous cases, $\alpha$
is almost constant across the gap, as shown in the upper right
panels of Figure \ref{fig:fig7}, and  the time and yz-direction averaged $\alpha$ 
is shown as the dotted curves in the 
lower left panel of Figure \ref{fig:fig12}.  For MHD disks with gap opening,
$\alpha_{Max}$  increases towards the gap region, as shown in the lower
right panel of Figure \ref{fig:fig7} and the averaged $\alpha_{Max}$ is shown as the solid curves in the
lower left panel of Figure \ref{fig:fig12}, again suggesting that the MHD stress 
 is  smoother across the gap than
the density or viscous stress. The time and yz-direction averaged magnetic field (net $B_{z}$) is also stronger
within the gap (the lower right panel of Figure \ref{fig:fig12}).
 
\begin{figure}
\epsscale{1.0} \plotone{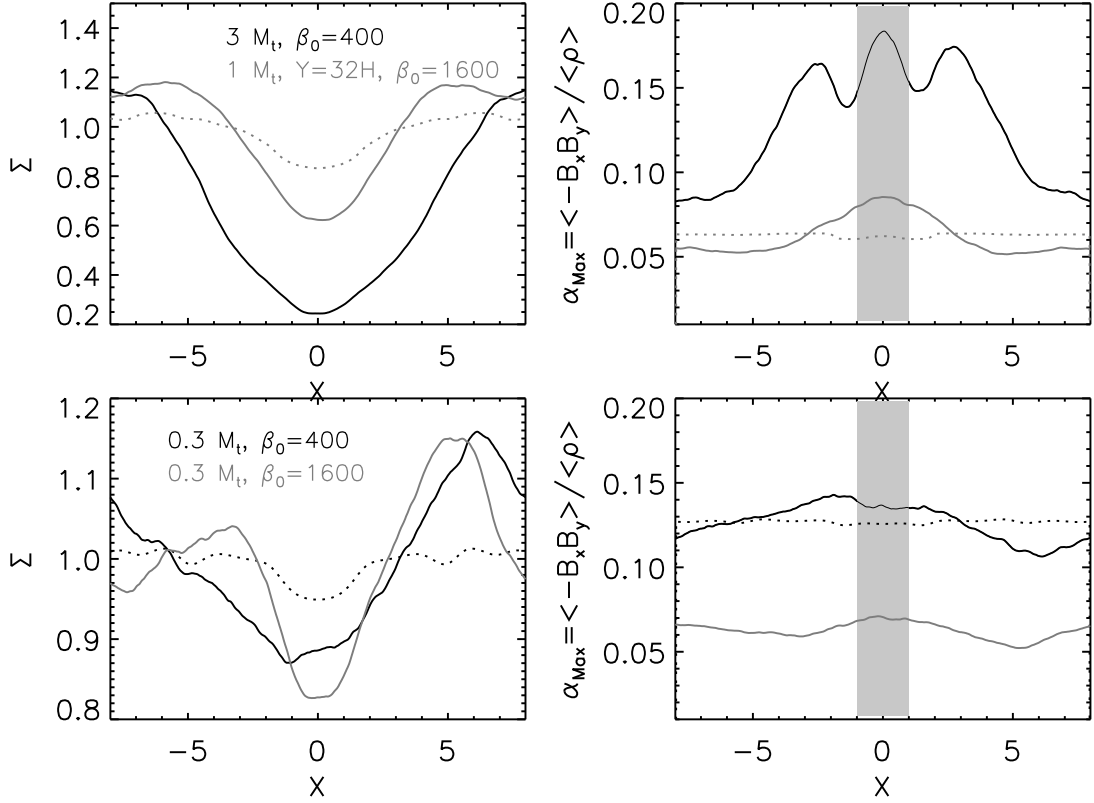} \caption{Similar to Figure \ref{fig:fig12} but for
more MHD and viscous cases. Upper panels: the case with a more massive planet 
 (M30B400), and a y-direction longer boxes (M10B1600b)).  The viscous case (V10bsv) is shown as the dotted curve.
 Bottom panels: the cases with a less massive planet and various net vertical field strengths (M03B400, M03B1600). The dotted curve is the viscous case V03. In all these cases, $\alpha$ of MRI disks peaks towards the gap region.
} \label{fig:fig9}
\end{figure}

\begin{figure}
\includegraphics[width=0.5\textwidth]{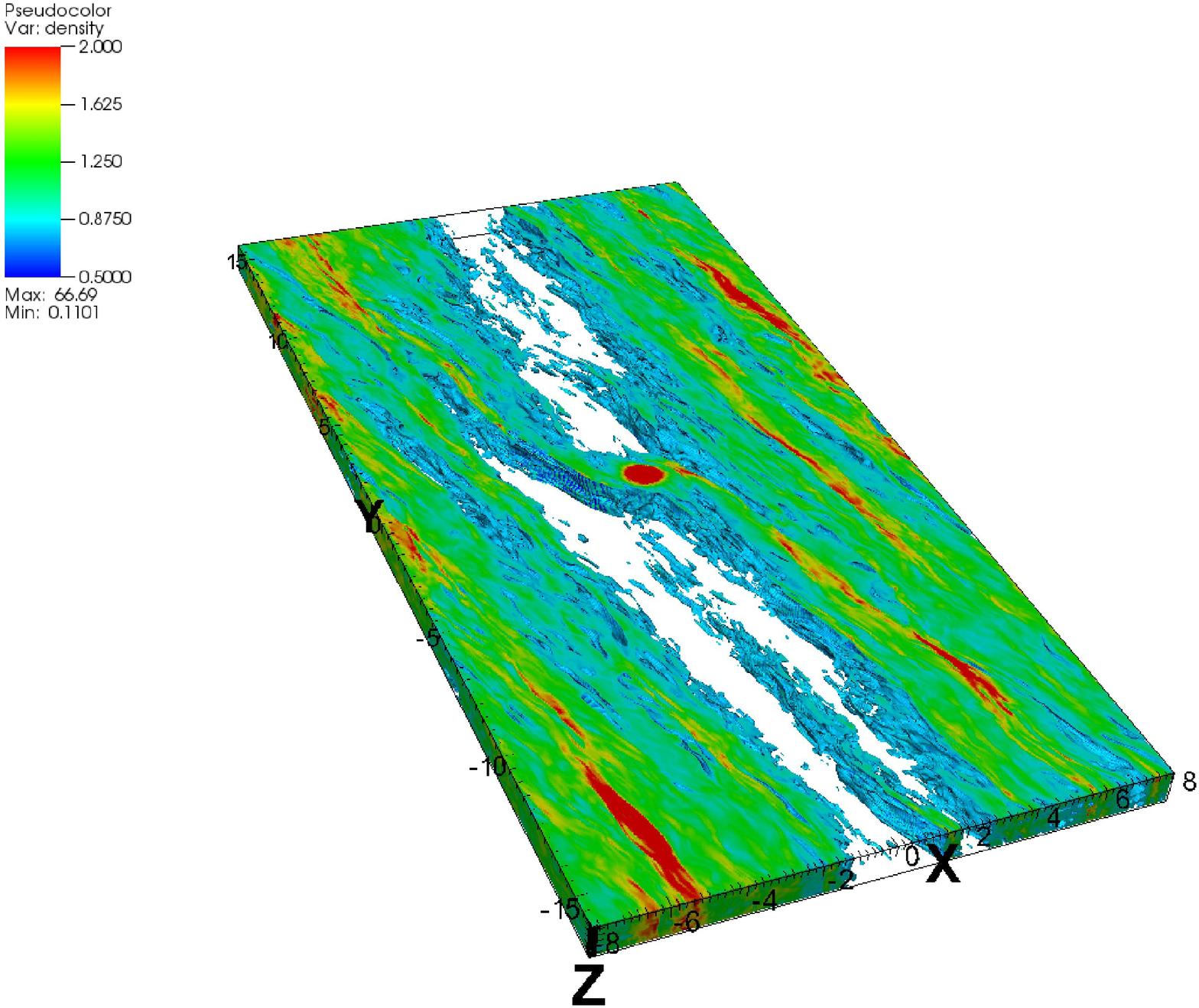} \hfil
\includegraphics[width=0.5\textwidth]{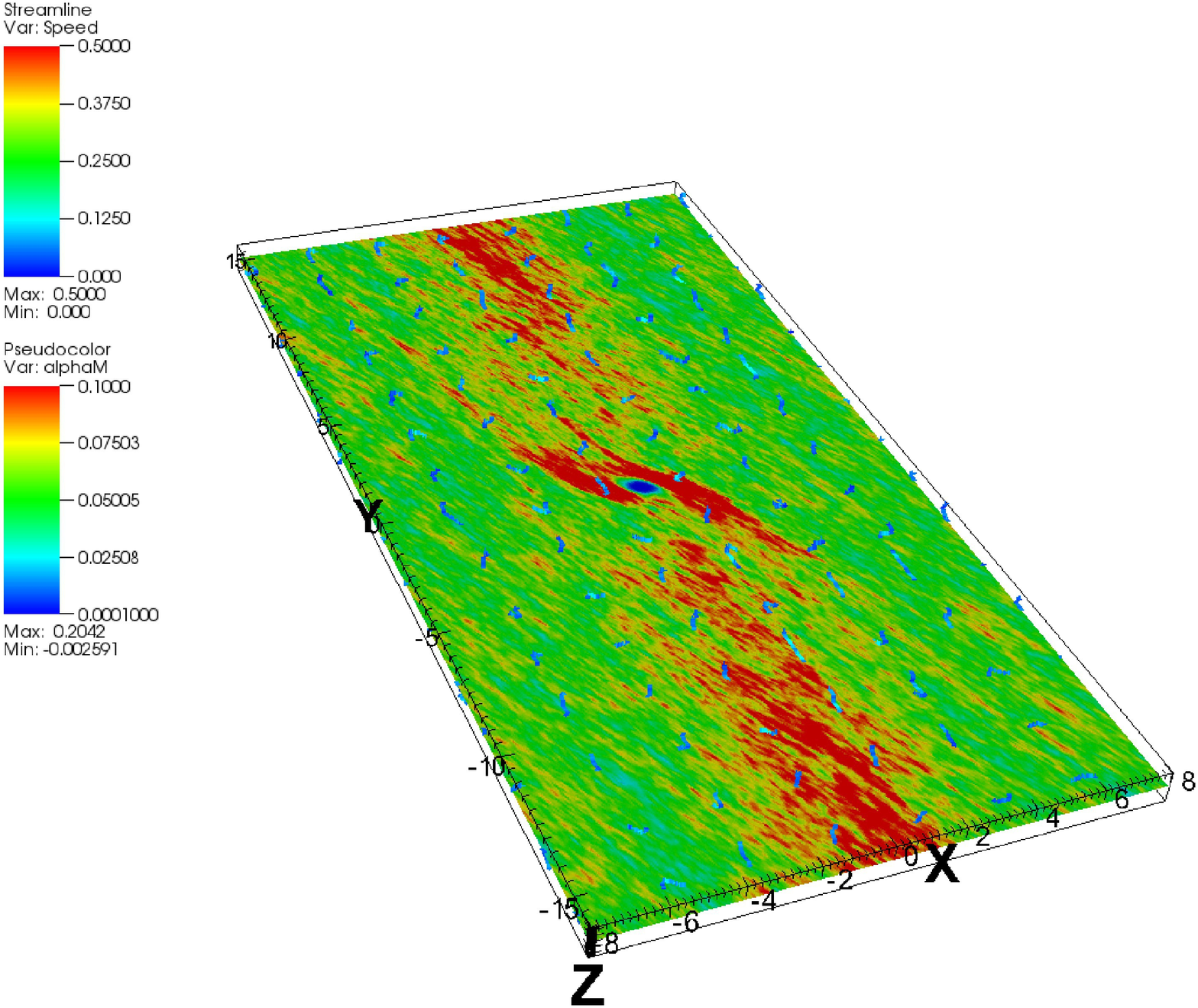} \\
\caption{ 
Left Panel: Volumetric rendering of the density during the interaction between the planet and MRI turbulent disks for M10B1600b. 
Regions with densities below 0.8 midplane density are removed.
Right Panel: The time averaged $\alpha_{Max}$ at the $z$=0 slice for 
this case.  The time averaged
magnetic field streamlines are also plotted. The turbulent component of the magnetic fields
is averaged out and the averaged fields show the net vertical field
geometry. 
The net vertical magnetic fields in M10B1600 seem to diffuse freely into the gap, causing
a higher $\alpha$. 
 }
\label{fig:fig17}
\end{figure}

We also explored cases with different planet masses (e.g. 3 M$_{t}$, 0.3 M$_{t}$),
and box sizes (e.g. 32 H in y direction), as shown in Figure \ref{fig:fig9}.
$\alpha_{Max}$ in these cases all show the same trend of increase towards the gap region, and
gaps are again deeper and wider in MRI disks than corresponding viscous cases. The density and 
$\alpha$ structure in the $xy$ plane for our biggest box simulation (M10B1600b) are shown in
Figure \ref{fig:fig17}, which nicely illustrate the turbulent nature of the disk and demonstrate the
higher $\alpha$ in the gap region.

\begin{figure}
\includegraphics[width=0.5\textwidth]{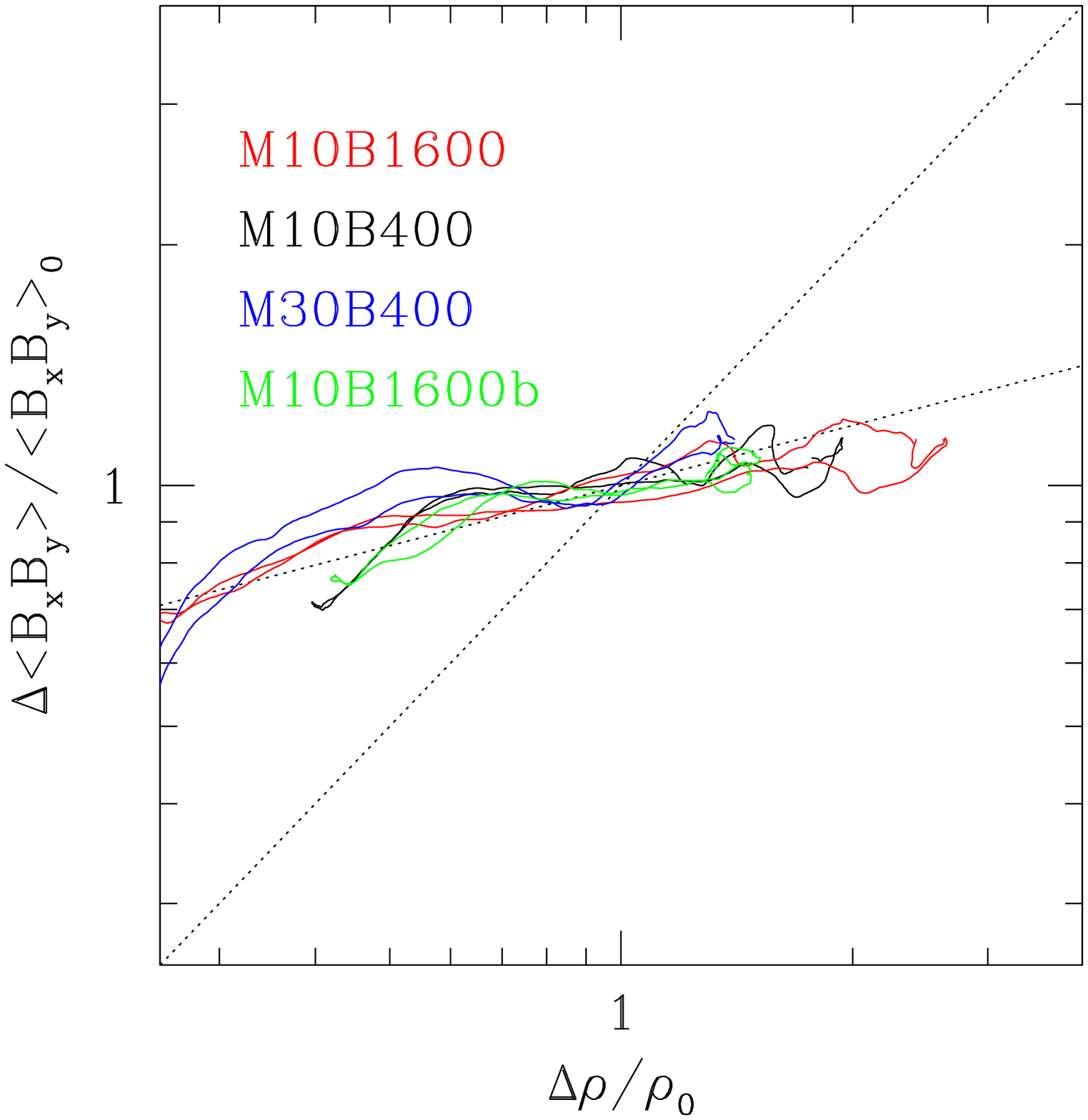} \hfil
\includegraphics[width=0.5\textwidth]{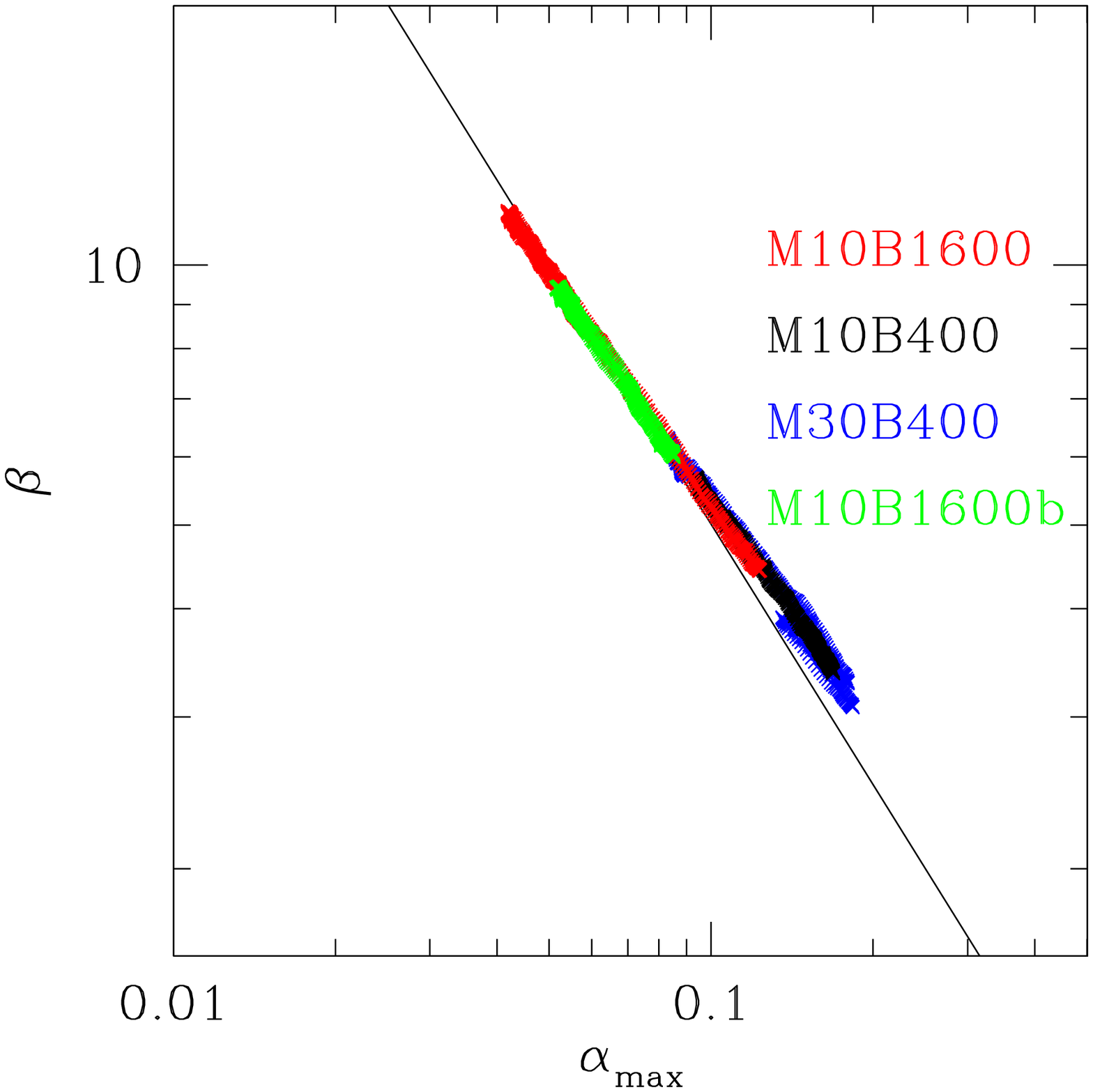} \\
\caption{Left panel: the excess of the $yz$-direction and time averaged 
Maxwell stress with respect to the excess of the averaged density at each x position across the gap region for the M10B400 (black curves),
M10B1600 (red curves), M30B400 (blue curves), and M10B1600b (green curves).  
$<B_{x}B_{y}>_{0}$ is the Maxwell stress at the position where the averaged $\rho=\rho_{0}=1$.
The dotted curve from the lower left to the upper right shows the stress-density 
relation in a viscous disk $\Delta T/T= \Delta \rho/\rho$, while the best fit for the stress across
the gap in MRI
disks is $\Delta T/T\propto 0.25\Delta \rho/\rho$.
Right panel: The $yz$-direction and time averaged $\beta$
with respect to the averaged disk
Maxwell stress $\alpha_{Max}$   at each x position
across the simulation domain. The point moves to the lower right when it crosses
the gap
 where it has lower $\beta$ and higher $\alpha_{Max}$. The solid line is  $\beta=1/2\alpha$.} \label{fig:fig10}
\end{figure}

In order to quantify this weak stress dependence on the density across the gap,
We calculate the deviation
of the Maxwell stress along $x$ 
\begin{equation}
\frac{\Delta \langle\langle B_{x}B_{y}\rangle\rangle}{\langle\langle B_{x}B_{y}\rangle\rangle_{0}}\equiv
\frac{\langle\langle B_{x}B_{y}\rangle\rangle_{x}-\langle\langle B_{x}B_{y}\rangle\rangle_{0}}{\langle\langle B_{x}B_{y}\rangle\rangle_{0}}\,
\end{equation}
with
respect to the deviation of the density ($\Delta \langle\langle\rho\rangle\rangle/\langle\langle\rho\rangle\rangle_{0}$), as shown
in the left panel of Figure \ref{fig:fig10}. Both $\langle\langle B_{x}B_{y} \rangle\rangle$ and
$\langle\langle\rho\rangle\rangle$ are derived from Figure \ref{fig:fig12} and Figure \ref{fig:fig9} and $\langle\langle B_{x}B_{y} \rangle\rangle_{0}$
corresponds to $\langle\langle\rho\rangle\rangle_{0}=1$. For the viscous stress
($T_{v}=1.5\nu\rho\Omega$), $\Delta T_{v}/T_{v}=\Delta \rho/\rho$.
However,  for all the MHD cases with net vertical magnetic flux, $\Delta \langle\langle B_{x}B_{y}\rangle\rangle/\langle\langle B_{x}B_{y}\rangle\rangle_{0}=C
\Delta \langle\langle\rho\rangle\rangle/\langle\langle\rho\rangle\rangle_{0}$ where $C\sim 0.25<1$, suggesting
$\langle\langle B_{x}B_{y}\rangle\rangle\propto \langle\langle\rho\rangle\rangle^{0.25}$ across the gap, and
indicating the weak dependence of the stress on the density across the gap. Thus, it is possible to
use this new relationship between stress and density to construct a viscous disk model to simulate the gap as in the MRI cases. However, we caution that this relationship is only
approximately true for these limited cases as shown.

Finally, in order to understand why $\alpha_{Max}$ is larger
inside the gap, we test if the empirical relationship between the turbulent
Maxwell stress $\alpha_{Max}$ and the plasma $\langle\langle\beta\rangle\rangle$ in MRI saturated states
($\alpha_{Max}\sim1/2\langle\langle\beta\rangle\rangle$) (Hawley \etal 1995) still stands across the gap
locally. This is shown in the right panel of Figure \ref{fig:fig10}, where each
point represents a $y$ direction averaged $\alpha_{Max}$ 
with respect to the averaged $\langle\langle\beta\rangle\rangle$ at
each radius $x$.  As $x$ marches towards the center of the gap, the
$\alpha_{Max}$ value increases and the $\langle\langle\beta\rangle\rangle$ value decreases. And they tightly following the empirical relationship. The
decreasing $\langle\langle\beta\rangle\rangle$ towards the gap
suggests that the turbulent magnetic fields are more concentrated in the gap region with respect to
the density.

Not only the turbulent magnetic fields, but also the net magnetic fields concentrate
to the gap region (the lower right panel of Figure \ref{fig:fig12}). These two evidences suggest efficient
global transport of magnetic fields to the low density gap region in MHD disks with net vertical
fields.

\subsection{Persistence of the gap}

As discussed in \S 4.2.1,
the effective $\alpha$ is higher in the gap region in net vertical flux MHD disks. Thus, 
if the gap depth is similar between MHD and viscous disks,
the Maxwell stress in net vertical flux MHD disks will have a smaller gradient than the viscous stress in viscous HD disks. Since it is the stress gradient that determines 
the global transport of the angular momentum and mass, any pre-existing density features will
behave quite differently among MHD disks and viscous HD disks. 

\begin{figure}
\epsscale{1.0} \plotone{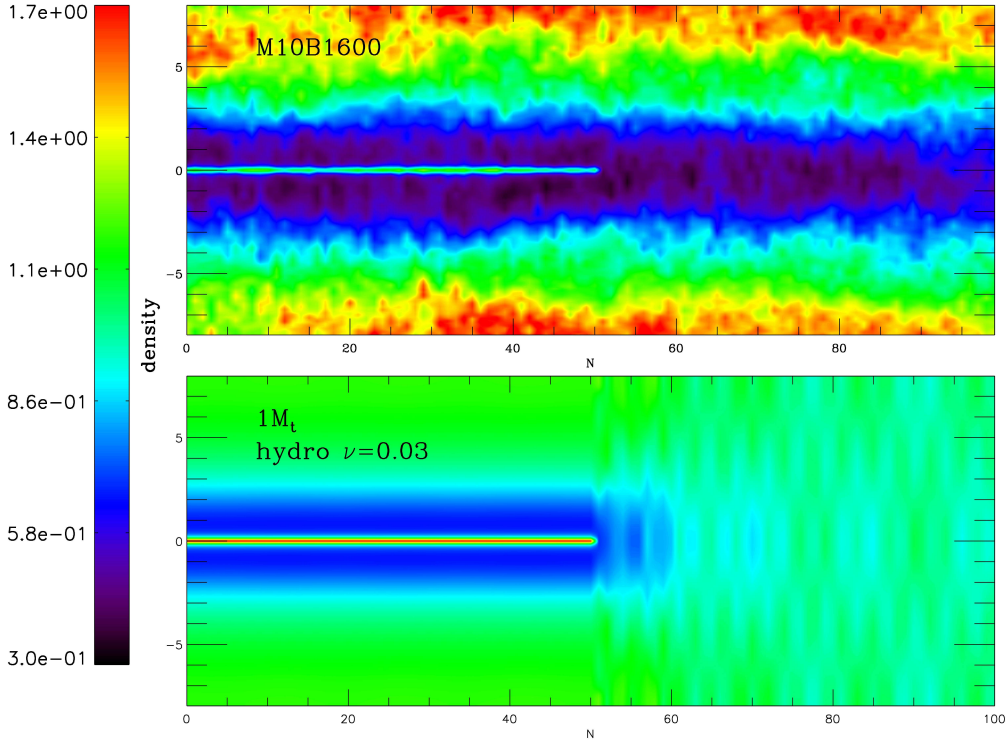} \caption{The space time plot for the $yz$-direction
averaged disk surface densities for the M10B1600 (net vertical flux), and V10la (viscous HD). However, the planet's
potential is suddenly changed to be zero in the middle of the simulation. As shown,
the gap feature in M10B1600 persists longer than that in a viscous disk. This is
 consistent with
the fact that the stress across the gap in MRI disks is more uniform than the density, so that
any existing density feature takes a longer time to be altered by turbulent stress in
net vertical flux MHD disks. 
} \label{fig:fig16}
\end{figure}

To test this hypothesis, we restart runs M10B1600 and V10l
at a time of 50 orbits, when gaps formed by the planet are fully developed, but with 
planetary potential switched off in the subsequent evolution.
The space-time plot for the disk surface
densities are shown in Figure \ref{fig:fig16}. As clearly shown in the upper panel, the
gap feature in M10B1600 even persists for 50 orbits  after the
planet disappears, while in viscous disks the gap is quickly closed
for $\sim$10 orbits comparable to the viscous timescale $(2
H)^{2}/\nu\sim20$ orbits. This confirms that the stress is uniform
in net vertical flux MRI disks, leading to the inefficiency in erasing the preexisting  density
features in the shearing box.

This result has far reaching implications for the measurement of the ``viscous" timescale
from the diffusion of density features in observed disks. If the disk is
threaded by a vertical magnetic field,
the disk evolution timescale will be longer than the ``viscous'' timescale. 

\section{Gap opening criteria}

\subsection{``Thermal Criterion''}
We have demonstrated that in an inviscid HD disk, a planet with
a mass 10 times smaller than the thermal mass can open gaps by shock
formation. Unlike gap opening by a massive planet ($\ge 1 M_{th}$)
where a single gap is opened around the planet, a low mass planet
opens two gaps away from the planet at the position where density
waves shock (Eq. \ref{eq:eqshock}).  

Simulations with a 0.3 $M_{th}$ planet embedded (M03B400, M03B1600)
start to show gap features around the planet (lower panels in Figure \ref{fig:fig9}).
Similar to 1 $M_{th}$ mass cases, a smaller  
$\alpha_{Max}$ leads to a deeper gap. 
We expect
a 0.3 $M_{th}$ planet can open a clearer gap in a even less turbulent disk.
Thus, we do not confirm the suggestion of 
Papaloizou \etal (2004) that the thermal
mass is a threshold for gap opening.  Low mass planets can
also open gaps as long as the disk stress is low. 

On the other hand, the timescale for gap opening is longer for
smaller planets, and eventually the feedback due to migration starts
playing an important role in the gap opening process(Ward 1997,
Rafikov 2002), which requires global simulations (Li \etal 2009).

In a future paper, we will show that 
there is also clear evidence that
the gap structure not only depends on the absolute value of the
stress ($\alpha$) but also on the field geometry. 

\subsection{``Viscous Criterion''}
The derivation of the traditional ``viscous criterion'' (Eq.
\ref{eq:viscous}) assumes the thermal criterion is satisfied
so that the disk scale height is replaced by the planet's Hill
radius.  If we relax this assumption, and balance the momentum flux
excited by the planet (Goldreich \& Tremaine 1980)
\begin{equation}
F_{H}\approx 0.93 (GM_{p})^{2}\frac{\Sigma_{0}R_{p}\Omega_{p}}{c_{s}^{3}}
\end{equation}
and the momentum flux due to viscosity
\begin{equation}
\dot{H_{\nu}}=3\pi \Sigma \nu R_{p}^{2}\Omega
\end{equation}
the viscous criterion becomes
\begin{equation}
\left(\frac{M_{p}}{M_{*}}\right)^{2}\gtrsim\frac{10 \nu H^{3}}{R_{p}^{5}\Omega}\sim 6\alpha\left(\frac{H}{R_{p}}\right)^5\,.
\end{equation}
where 6 is just an order of magnitude estimate and the detailed value depends on
how we define a ``gap''.
However, this criterion needs to be modified to study gap opening in the shearing box limit, since
the momentum flux due to viscosity in the shearing box is
\begin{equation}
\dot{H_{\nu}}=1.5 \Sigma \nu R_{p}\Omega H_{y}
\end{equation}
where $H_{y}$ is the box size in $y$ direction. Thus, the viscous criterion in the shearing box is
\begin{equation}
\left(\frac{M_{p}}{M_{*}}\right)^{2}\ge\frac{1.6 \nu H_{y}H^{3}}{R_{p}^{6}\Omega}\sim\alpha \left(\frac{H}{R_{p}}\right)^5\frac{H_{y}}{R_{p}}\,,
\end{equation}
or
\begin{equation}
\left(\frac{M_{p}}{M_{th}}\right)^{2}\ge\frac{1.6 \nu H_{y}\Omega^{2}}{c_{s}^{3}}\sim \alpha \frac{H_{y}}{H}
\end{equation}

For a disk having $\nu$=0.113, the viscous criterion requires the planet mass larger than 
1.7 $M_{th}$ with $H_{y}=16 H$. This explains why a gap cannot be opened  in our HD viscous runs ($10\%$ dip in V10).
However, a $50\%$ deep gap is opened in our MHD run (M10B400) as discussed in \S 4.2, suggesting that 
the viscous criterion needs
to be modified, at least for net vertical flux simulations. 

\subsection{The Comparison with Previous results}
Our results are different from previous gap opening studies using net toroidal flux
or zero flux MHD simulations  (Winters \etal 2003,  Nelson \& Papaloizou 2003, Papaloizou
\etal 2004, Uribe \etal 2011) in respect that previous simulations found similar gap depths between MHD cases
and viscous cases \footnote{ There are some subtle differences in these previous studies.
 Winters \etal (2003) found gaps are slightly shallower in MRI disks compared with viscous cases,
while Nelson \& Papaloizou (2003) suggest that
gaps are deeper and wider in MRI disks.}. Furthermore, our main result that  $\alpha_{max}$ is higher in the whole gap region 
has not been reported before. Papaloizou \etal (2004) reported the increasing $\alpha_{max}$ along the 
planetary wake in their Figure 19 \& 21. But their Fig 21 suggests that $\alpha_{max}$ decreases 
in the gap region. 
These differences between our results and previous results may suggest that different magnetic
field geometries in the disk affect the gap opening process.

To confirm that the deeper gaps in our simulations are due to the different field geometry
we have applied, we have also carried out gap opening simulations with net toroidal magnetic fields. 
We have indeed observed similar planet-induced
gap depths between net toroidal flux MHD cases and viscous cases, which is consistent
with previous studies. 

More surprisingly, we found that, in net toroidal flux MHD simulations, $\alpha_{max}$
decreases towards the gap region (consistent with Figure 21 in Papaloizou \etal 2004), which
is fundamentally different from the increasing $\alpha_{max}$ towards the gap region in net vertical flux MHD simulations. 
This result will be presented in a later paper. 

\section{Discussion}

\subsection{Torque and Migration}

Planet migration is an important result of the 
planet-disk interaction theory (Goldreich \& Tremaine 1980, Lin 
\& Papaloizou 1993).  As we have discussed in \S 4.1.3,
interaction with a MHD turbulent disk changes the planetary torque
density in comparison to HD cases.  More importantly, it causes
a ``random walk'' of the planet when fluctuations of the torque
associated with turbulence dominates the Lindblad torque (Nelson
\& Papaloizou 2004, Uribe \etal 2011).

\begin{figure}
\includegraphics[width=0.3\textwidth]{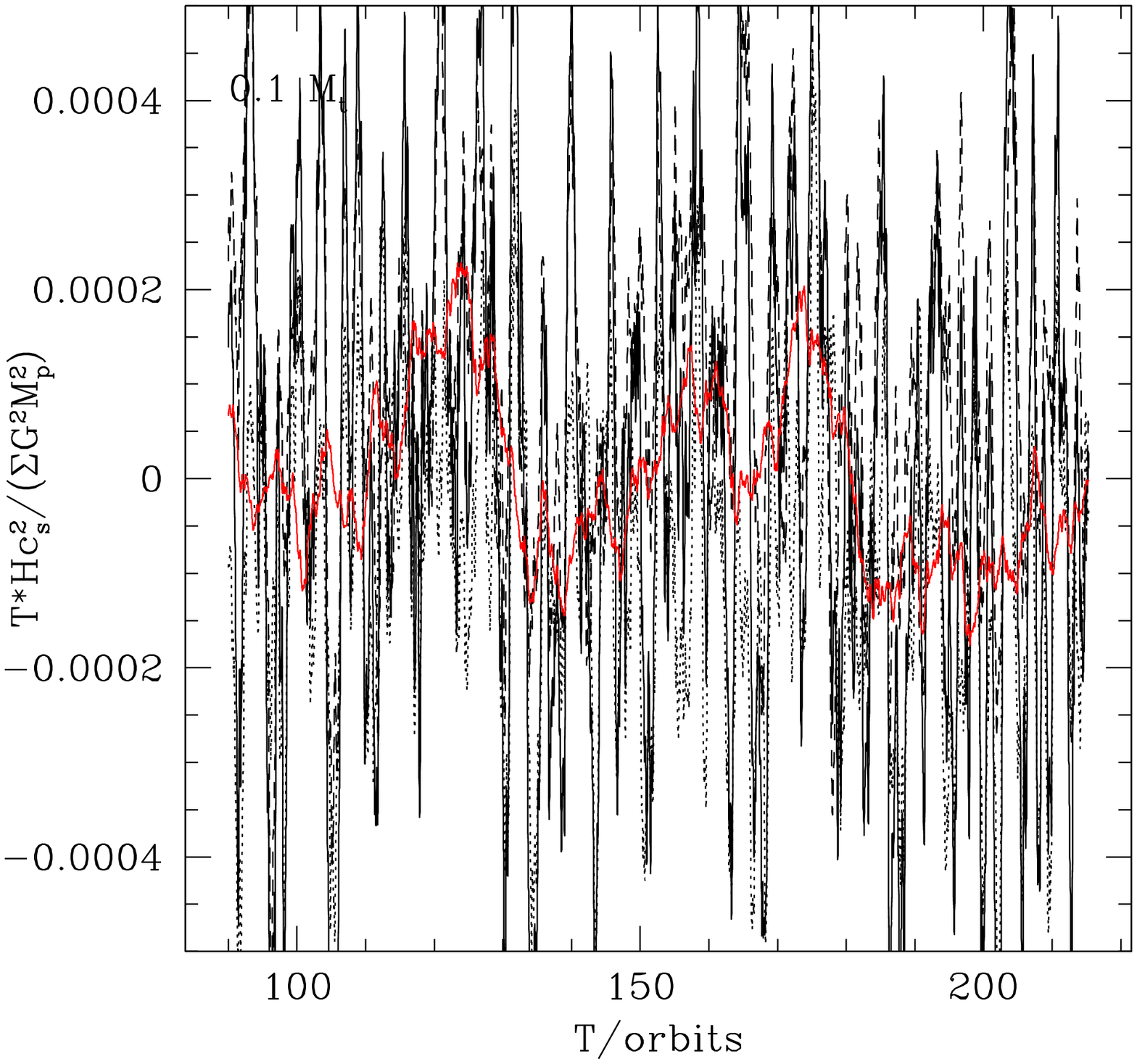} \hfil
\includegraphics[width=0.3\textwidth]{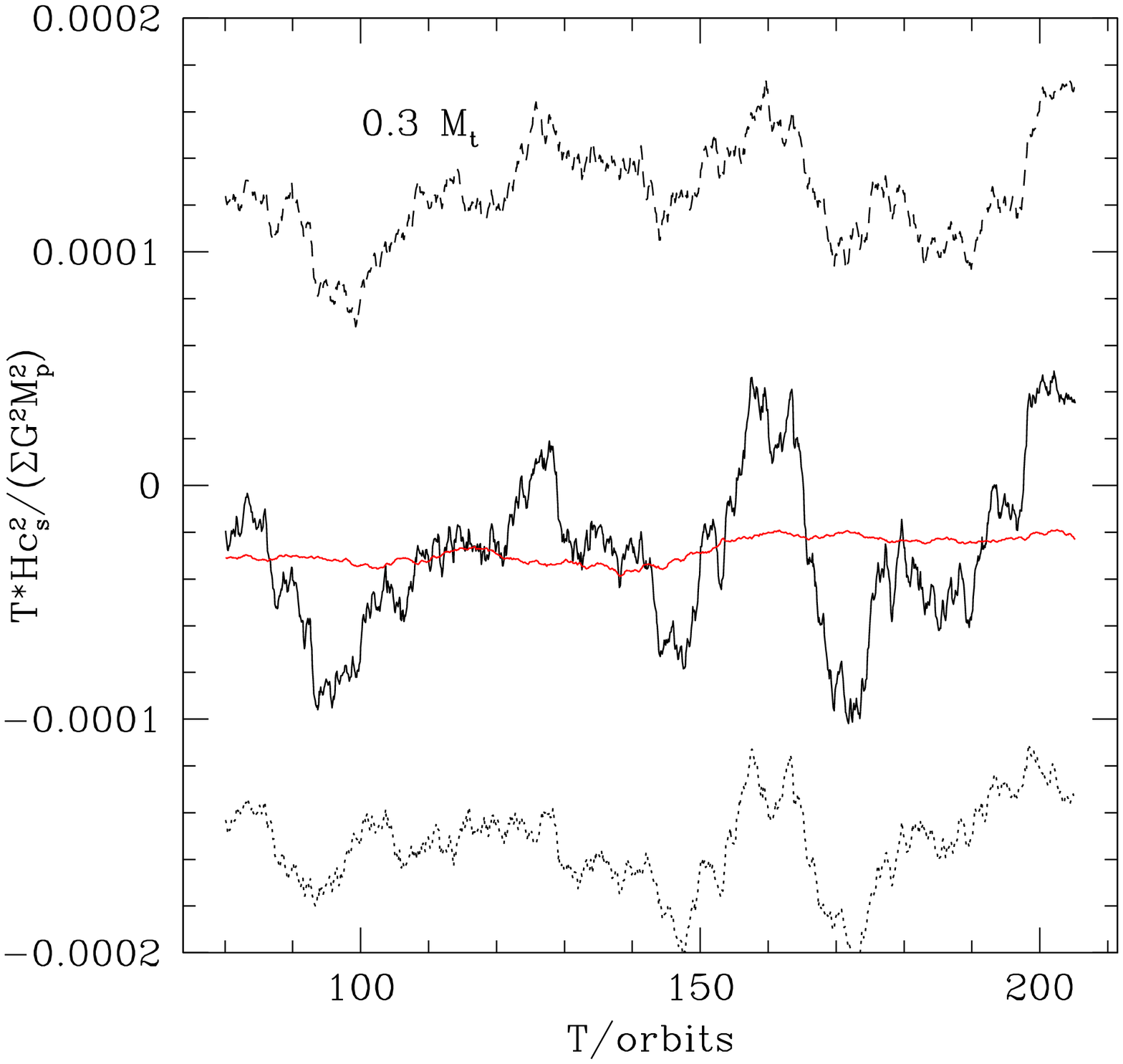} \hfil
\includegraphics[width=0.3\textwidth]{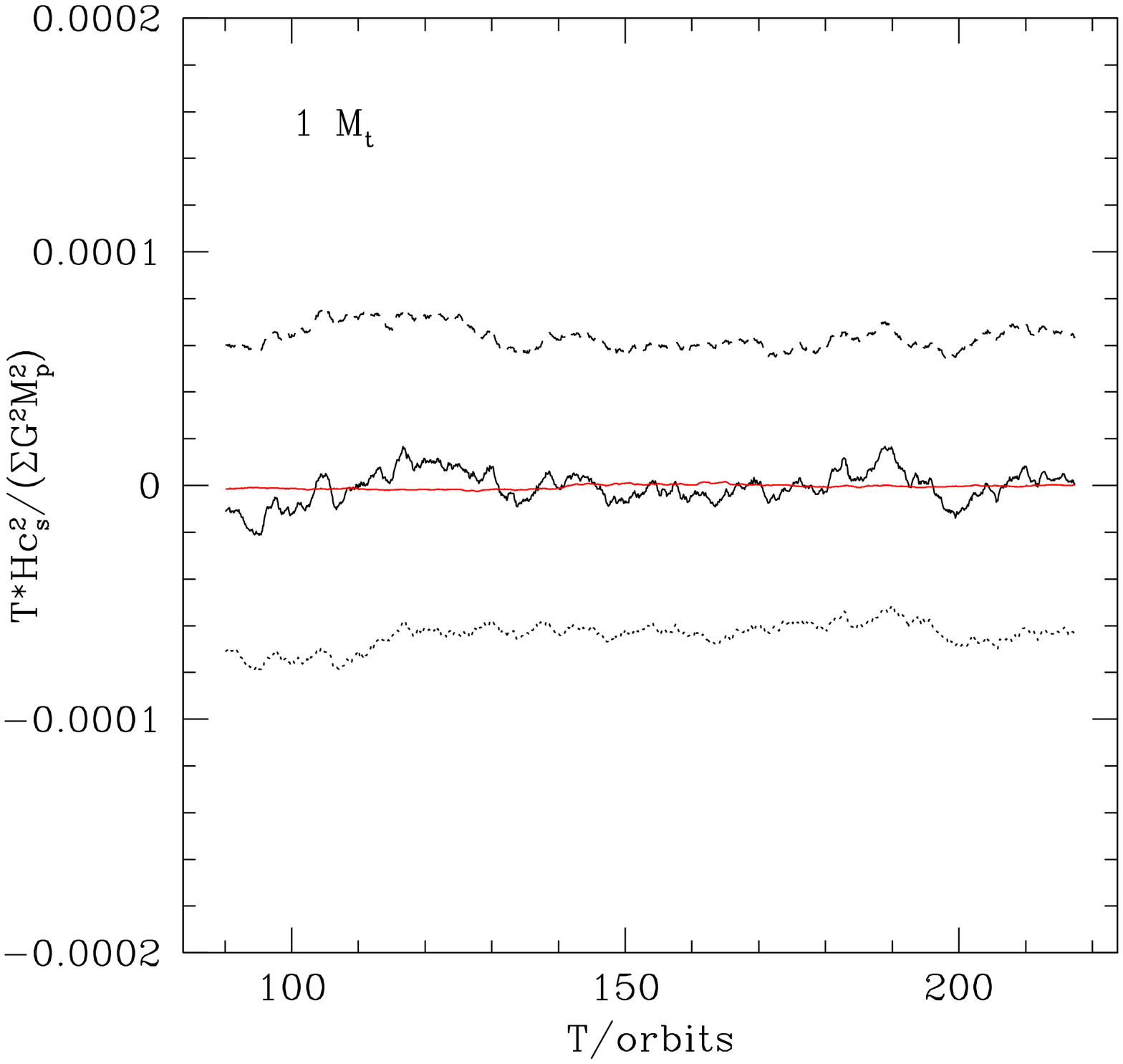} \\
\caption{The integrated torques with time for M01B400 (1 $M_{th}$), M03B400 (0.3 $M_{th}$), M10B400 (0.1 $M_{th}$) (left to right panels).
The dashed curves are the torques from $x>0$ side of the disk while the dotted curves are from $x<0$
side of the disk. The solid curves are the net torque. These curves are averaged over 1 orbit period.
The red curves are the net torque
averaged over 10 orbits period. For less massive planets, the torque due to the turbulence can dominate the Lindblad torque, leading to the planet's random walk.
For the intermediate mass planet (0.3 M$_{t}$ case), the net torque is actually non-zero (negative in this case), contradictory to the expectation that, due to the symmetry of the shearing box, the averaged net torque should be zero.
This non-zero net torque is caused by the non-uniform disk background from the large scale MRI zonal flow. }
\label{fig:figtorque}
\end{figure}

The time evolution of the total torques for runs M01B400, M03B400,
and M10B400 are shown in Figure  \ref{fig:figtorque}.  The torques
from both sides of the disk and the net torques are shown.  The
fluctuating torque is more important for less massive planets (e.g. 
0.1 M$_{t}$, M01B400)
since
the torque from the turbulence is proportional to the planet mass
while the Lindblad torque is proportional to the square of the
planet mass.

For an intermediate mass planet (0.3 M$_{t}$, M03B400), 
the long lasting non-uniform background
zonal flow (Johansen \etal 2009, Simon \etal 2012) can also significantly change the
migration (Yang \etal 2009, 2012).  This is shown in the middle panel of Figure \ref{fig:figtorque}
where the net torque is always negative due to the zonal flow, even though 
it should be zero if the background is uniform in the shearing box limit.  This net torque
is almost 20$\%$ of the total one-sided torque. However, caution has to be made
by using shearing box simulations to study both stochastic migration and
zonal flows (Yang \etal 2009,2012), and how zonal
flow affects planet migration in global disks needs further study.

It is unclear whether, in a very long time scale, the planet's stochastic motion can still lead to a 
one direction migration as in a laminar disk.  We can indirectly address this question
by averaging the one-sided torque over long times and comparing it with the
Lindblad torque in HD disks.  As shown in Figure \ref{fig:torquet},
the averaged torques in net vertical flux MHD disks do not equal to the torque 
in HD disks. An excess torque very close to the planet is present around 0.2-0.3 H. A similar excess torque is
found by Baruteau \etal (2011) (Figure 13) in global simulations. 

Several mechanisms can potentially explain this excess torque, such as
the ``magnetic resonance" (Terquem 2003, Fromang \etal 2005) and the horseshoe motion amplified
corotation torque (Guilet \etal 2013). Unfortunately, we did not confirm any of these mechanisms, partly due 
to the complication of MHD turbulence.
For the ``magnetic resonance'', We do not observe significantly density features associated with this resonance
 (Fig. 4 in Fromang \etal 2005). For the MHD corotation torque, our
shearing box set-up do not exhibit the shift between the planet and the separatrices by design.
This shift is global disks introduces the MHD corotation torque (Guilet \etal 2013). Thus,
the mechanism behind this excess torque in our simulations is unclear.
Examining Fig. 3 in detail, we find that the circumplanetary region
in MHD cases is asymmetric which may indicate some modification by the magnetic field
to the circumplanetary region.

\subsection{Can we use $\alpha$ to study gap opening?}

Although viscous disks have been extensively used to study gap
opening and even planet accretion, our work suggests that caution
has to be exercised when applying viscous models to real protoplanetary
disks.

First, $\alpha$ viscosity models MRI turbulence only in a statistical
sense (Balbus \& Papaloizou 1999).  At any given instant in time,
quantities such as the wake profile, gap shape, and torque can be
very different in inviscid MHD turbulence compared to a viscous HD
disk. This implies that if we are able to resolve the spiral structure
induced by planets in protoplanetary disk with ALMA, the spiral arm may be distorted  and disjointed
which is very different from the smooth structure 
in viscous HD disks (Fig \ref{fig:fig3}).

Second, even though the stress in MRI turbulent disks can be scaled
to the disk pressure in a volume averaged sense, we find that since
the turbulent stresses are not proportional to density, possibly due to global
transport of magnetic fields, the value
of the effective $\alpha$ can vary across the gap region.  In our net 
vertical flux MHD
simulations, the gap region has a higher $\alpha$ than the rest of
the disk.  
Calculating accurate models in viscous HD requires varying
the value of $\alpha$. An empirical formulae is presented in \S 4.2.1. 
Note that this formulae is derived from our simulations with limited
net field strength range, planet mass range, and ignoring dissipation.
Particularly, the MHD turbulent strength can be affected by the dissipation coefficients 
(Lesur \& Longaretti 2007, Longaretti \& Lesur 2010) even in net vertical field simulations.
In order to understand the gap structure in the realistic simulation with
dissipation, we need to first understand how magnetic fields are transported
in such disks with density features. This demands numerical simulations with
very high resolution, which is beyond the scope of this paper.

Finally, properties of the circumplanetary region highly depend on
the magnetic field geometry and field strength there, which is not
isolated from the rest of the magnetic fields in MHD disks. Although 
the density around the circumplanetary region is very high due to the gravity
of the planet, the magnetic fields could be less affected by the presence
of the planet due to the efficient global transport of the magnetic fields. 
If $\alpha\sim 1/2\beta$
still stands there, the equivalent
$\alpha$ in the cicumplanetary region could be very small.  Moreover, the
global geometry of the magnetic field in this region means the
stress is not completely local, but there are torques due to
connections to the outer gap regions.  It is unclear how this can
be modeled in the local $\alpha$ formulation.

\subsection{Gaps in protoplanetary disks and observational Implications}
Current observations cannot constrain the structure of protoplanetary
disks accurately due to the limited resolution and disk's large optical depth at infrared. Theoretical
calculations considering ionization network (e.g. Bai 2011) suggest the existence
of the MRI ``dead zone'' (Gammie 1996). The stress (or $\alpha$)
can be very low in the ``dead zone'' making it similar to the
inviscid HD disk.  Based on our inviscid HD simulations, two
gaps can be opened in the disk adjacent to a low mass planet. This
``two gap'' structure might be observable by ALMA in future.

For MRI active regions in disks, the planet-induced gap structure depends on
how the disk is threaded by global magnetic fields.  Disks with net vertical 
fields have drawn greater attention recently since net vertical fields seem to
be essential to produce disk winds and jets 
(Bai\&Stone 2012, Fromang \etal 2012, Lesur \etal 2012).
If protoplanetary disks are threaded by net vertical fields, gaps can be
opened by lower mass planets in contrast to the viscous HD models.

On the other hand, if a gap is
observed, ``$\alpha$ models'' may not be
accurate enough to use the gap structure to determine the planet mass. 
Stratified MHD simulations with 
realistic disk ionization structures are needed. 
Eventually global
MHD simulations are needed to derive a realistic gap shape.
Some of these issues will be addressed in our next paper.

Furthermore, the spiral arms produced by the planet can be fragmented
by MRI turbulence. 
In a real protoplanetary disk, the spiral arms can be turbulent and dynamical,
and different from the static spiral arms produced in viscous ``$\alpha$ models''.

\section{Conclusion}

We have studied the wake and gap opening by a low mass planet with
both two dimensional HD and three dimensional MHD simulations in
the unstratified local shearing box limit.  Similar to the case of
inviscid HD,  the density wake launched by a 
planet steepens into a shock in the MRI turbulent disk. Thus shock dissipation dominates wave
dissipation in MHD disks. Turbulent fluctuations cause the shock position
to vary in time, and when averaged over hundreds of orbits the random
wake positions form a smooth density profile which is remarkably similar
to that in viscous disks with the same total stress.  On the other
hand, we find the average torque density excited
by the planet to be different from the torque density in HD
simulations: it has a peak around 0.2-0.3
H away from the planet compared to
$\sim$1 H in HD simulations. Furthermore, zonal flow
makes the background disk not uniform leading to a net torque even
in the local shearing box limit.

We have also studied gap opening criteria in both HD disks and
 MRI-driven MHD turbulent disks.  In inviscid
HD disks, we find a 0.1 thermal mass planet can open two gaps at
the position where the density waves shock, which subsequently deepen and
merge into a single gap structure. This is  consistent
with the expectation from nonlinear semi-analytical calculations
by Goodman \& Rafikov (2001) and Rafikov (2002b), but contradicts the generally accepted
``thermal criterion" for gap opening.   For a disk with H/R=0.1, the
timescale for this gap opening is around $10^{3}$ orbits.  In
MHD turbulent disks, we find that a 0.3 $M_{th}$ planet can  open a $20\%$
deep gap in a disk with equivalent $\alpha$=0.08.  
Overall, both our HD and MHD results show that gaps can be
opened by planets significantly below the thermal mass.

By varying the strength of the initial vertical magnetic fields, we have
studied the gap structure in MRI disks with different turbulent
amplitudes, and compared the results to viscous HD disks with the same
stress.  The gaps in MRI disks having net vertical fields are deeper and wider
in comparison to viscous HD disks.  In particular, we find that, with an initial
vertical field $\beta_{0}=400$ in disks, the gap formed by a thermal
mass planet contains of a density dip of 2, while only a 10$\%$
density dip is observed in the viscous HD disk. This
difference in gap shape is due to 
the fact that the MRI stress ($T$)
across the gap depends only weakly on the density ($\Delta T/T=
0.25 \Delta \rho/\rho$ in disks with initial vertical magnetic
fields) while the viscous stress is proportional to the density.

Looking at this in a different way, the equivalent $\alpha$ parameter of the
MHD turbulence is higher within the planet-induced gap of net vertical flux MHD disks.
Remarkably, the correlation between the total stress
measured by $\alpha$ and the plasma $\beta$ ($\sim 1/2 \alpha$)
found generally in studies of the MRI (Hawley \etal 1995) still holds even inside
the gap. Both turbulent magnetic fields and net vertical magnetic fields are
more concentrated in the gap region with respect to the density, suggesting
an efficient global transport of magnetic fields into the low density gap which
leads to $\alpha$ variation across the gap.

The fact that $\alpha$ is not constant within the gap, and that MHD
stresses are not proportional to density across the gap (leading to deeper and
wider gaps in MHD) calls into question the use of viscous HD simulations
to study gap opening in protoplanetary disks.  In the future, in order to model
the dynamics of protoplanetary disks more accurately, we
will study gap opening in stratified and layered MRI-turbulent disks.

\acknowledgments
This work was supported by NSF grant AST-0908269 and Princeton University.
All simulations were carried out using computers supported by the
Princeton Institute of Computational Science and Engineering.
We thank Steve Lubow and Xuening Bai for
helpful discussions.

\begin{table}
\begin{center}
\caption{Hydrodynamic Models (2D) \label{tab1}}
\begin{tabular}{ccccc}

\tableline\tableline
Case name &  Box size & Planet mass & Run Time & Kinematic  \\
                      &     H$\times$H              & M$_{t}$        & 2$\pi$/$\Omega$& Viscosity \\
\tableline
Inviscid Disks\\
\tableline
I1  & 8$\times$16 & 0.1 & 400 & 0  \\
I2  & 16$\times$16 & 0.1 & 400 & 0  \\
I3  & 32$\times$16 & 0.1 & 400 & 0  \\
I10inv  & 16$\times$16 & 1 & 20 & 0   \\
I03inv & 16$\times$16 & 0.3 & 20 & 0  \\
I01inv  & 16$\times$16 & 0.1 & 20 & 0  \\
 \tableline
Viscous Disks &&&&\\
\tableline
V01  & 16$\times$16 & 0.1 & 100 & 0.113  \\
V03  & 16$\times$16 & 0.3 & 100 & 0.113  \\
V10  & 16$\times$16 & 1 & 100 & 0.113  \\
V10l  & 16$\times$16 & 1 & 100 & 0.03  \\
V10sv  & 16$\times$16 & 1 & 100 & 0.056  \\
V10bsv  & 16$\times$32 & 1 & 100 & 0.056  \\
\tableline
\end{tabular}
\end{center}
\end{table}

\FloatBarrier

\begin{table}
\begin{center}
\caption{MHD Models (unstratified 3D (X=16H, Z=1H))\label{tab2}}
\begin{tabular}{cccccc}

\tableline\tableline
Case name &   Planet  & Run Time & Net Field  & Initial Field &stress  \\
                      &     mass (M$_{t}$)        & 2$\pi$/$\Omega$ &Geometry & $\beta_{0}$& $\langle\alpha\rangle$\\
\tableline
Y=16H\\
\tableline
B400 &  0 & 200 &  Vertical & 400 & 0.18 \\
M01B400 &  0.1 & 400 & Vertical &400 & 0.17\\
M03B400 &  0.3 & 246 &  Vertical &400 & 0.17\\
M03B1600 &  0.3 & 264 &  Vertical &1600 & 0.085 \\
M10B400 &  1 & 258 & Vertical &400 & 0.17\\
M10B1600 &  1 & 225 &  Vertical &1600 & 0.081\\
M30B400 &  3 & 258 &  Vertical &400  & 0.12\\
\tableline
Y=32H\\
\tableline
M10B1600b &  1 & 138 &  Vertical &1600 & 0.084\\
\tableline
\end{tabular}
\end{center}
\end{table}

\clearpage

\end{document}